# Ring-originated anisotropy of local structural ordering in amorphous and crystalline silicon dioxide

Motoki Shiga[1,2,3,*,†], Akihiko Hirata[4,5,6], Yohei Onodera[6,7], Hirokazu Masai[8]

*Corresponding author
†motoki.shiga.b4@tohoku.ac.jp

[1] Unprecedented-scale Data Analytics Center, Tohoku University, 468-1 Aoba, Aramaki-Aza, Aoba-ku, Sendai 980-8578, Japan

[2] Graduate School of Information Science, Tohoku University, 6-3-09 Aoba, Aramaki-aza Aoba-ku, Sendai, 980-8579, Japan

[3] RIKEN Center for Advanced Intelligence Project, 1-4-1 Nihonbashi, Chuo-ku, Tokyo 103-0027, Japan

[4] Department of Materials Science, Waseda University, 3-4-1 Ohkubo, Shinjuku, Tokyo 169-8555, Japan.

[5] Kagami Memorial Research Institute for Materials Science and Technology, Waseda University, 2-8-26 Nishiwaseda, Shinjuku, Tokyo 169-0051, Japan

[6] Center for Basic Research on Materials, National Institute for Materials Science, 1-2-1, Sengen, Tsukuba, Ibaraki 305-0047, Japan

[7] Institute for Integrated Radiation and Nuclear Science, Kyoto University, 2-1010 Asashiro-nishi, Kumatori-cho, Sennan-gun, Osaka 590-0494, Japan

[8] National Institute of Advanced Industrial Science and Technology, 1-8-31 Midorigaoka, Ikeda, Osaka 563-8577, Japan

## Abstract

Rings comprising chemically bonded atoms are essential topological motifs for the structural ordering of network-forming materials. Quantification of such larger motifs beyond short-range pair correlation is essential for understanding the linkages between the orderings and macroscopic behaviors. Here, we propose two quantitative analysis methods based on rings. The first method quantifies rings by two geometric indicators: roundness and roughness. These indicators reveal the linkages between highly symmetric rings and crystal symmetry in silica and that the structure of amorphous silica mainly consists of distorted rings. The second method quantifies a spatial correlation function that describes three-dimensional atomic densities around rings. A comparative analysis among the functions for different degrees of ring symmetries reveals that symmetric rings contribute to the local structural order in amorphous silica. Another analysis of amorphous models with different orderings reveals anisotropy of the local structural ordering around rings; this contributes to building the intermediate-range ordering.

**Introduction**

Amorphous materials exhibit a disordered structure owing to the lack of translational periodicity. Unlike crystals, which exhibit translational periodicity, amorphous materials do not have a long-range structural order. The formation of a short-range order, whose definition is based on the distances between two nearest neighbor atoms, can be observed in the pair distribution function (PDF) obtained via X-ray diffraction (XRD) and neutron diffraction (ND) measurements for both crystals and amorphous materials. For example, the short-range order of amorphous and crystalline silica ($SiO_2$) under ambient conditions is based on a regular $SiO_4$ tetrahedron. The interconnection of tetrahedra with corner-sharing oxygen atoms forms a network structure. Meanwhile, for some amorphous materials, such as $SiO_2$, these diffraction experiments also provide the evidence of a structural order on a length scale larger than the atomic bond length, i.e., the intermediate range. The formation of intermediate-range order of amorphous $SiO_2$ is often discussed in terms of the first sharp diffraction peak (FSDP) observed in the diffraction data[1,2,3,4]. However, the origin of FSDP[3,5,6,7,8] in amorphous $SiO_2$ has long been debated. It is currently established that the length scale of FSDP is approximately 4 Å $\left(= \frac{2\pi}{q_{FSDP}}\right)$, $q =$ $\left(\frac{4\pi}{\lambda}\right) \sin\theta$ with the scattering angle $2\theta$ and X-ray or neutron wavelength $\lambda$ and that the periodicity with the coherence length is approximately 10 Å $\left(= \frac{2\pi}{\Delta q_{\text{FSDP}}}\right)$, where $\Delta q_{\text{FSDP}}$ is the width of FSDP[9]. Structural order analysis based on PDF has limitations caused by a rapid decrease in the atomic pair correlation peak intensity on the intermediate-range scale. Moreover, structural units of intermediate-range orders are larger complexes than chemical bonds, and they cannot be identified using pairwise correlation analysis. Hence, revealing the structural orders hidden in many-body correlations in amorphous materials remains challenging.

Structural models of glass, based on a network structure consisting of corner-sharing tetrahedral motifs, as found in $SiO_2$, help to investigate the intermediate-range structural orders in amorphous materials. The first structural model for glass was the crystallite model proposed by Frankenheim[10] in 1835, which described glass as an aggregate of small crystallites. The crystallite model, also known as the discrete crystalline model, presumes that crystalline clusters are embedded in a glass matrix, owing to the relatively sharper feature of FSDP in the diffraction data. However, this model has been found inappropriate because the estimated lattice constants of the crystallites are larger than those of $\beta$-cristobalite, thereby leading to a discrepancy between the observed and calculated densities. Subsequently, continuous crystal[11] and quasi-crystal models[7] have been proposed. Meanwhile, a continuous random network model was first proposed by Zachariasen[12] in 1932, wherein short-range structural units in glass were connected randomly. Zachariasen also addressed the relationship between the structure and glass-forming ability and proposed four rules for glass formation. Although Zachariasen's models are only based on the short-range-order characteristics, such as chemical bonds and the coordination number, such models have been widely used. Several structural models such as the layer model[13] and periodic boundaries of void (or cage) models[8, 14], which assume larger structural units, have been developed further. To validate such model assumptions on the structural order for various materials, statistical analysis methods using geometric and/or chemical structure information are useful. A major approach for characterizing the structural order is ring (closed-path) analysis in a chemically bonded network generated from a structural model. In the conventional analysis, rings are exhaustively enumerated by the shortest-path algorithms, and ring size distributions are analyzed[15,16,17,18,19]. Recently, Onodera et al.[20] reported structural orderings related to ring transformation in densified silica glass. However, since there is no analytical method for ring geometry, the discussion is limited to the number of atoms constituting the ring. For deeply understanding the intermediate-range order in amorphous materials, various analysis methods for the geometric information of rings and their linkages to structural order are necessary. This progress on understanding intermediate-range structural orders entails uncovering the relationship between the orders and material properties using quantitative analysis methods[21,22], which leads to a new path for designing novel functional materials. Furthermore, such methods contribute to building data-driven material design using large-scale structural models of disordered materials[23,24] generated by computational simulations using machine learning potentials[24,25,26,27,28], which can realize much faster implementation without compromising the theoretical accuracy compared to exact computations based on the density functional theory.

Here, we propose extensions from conventional ring analysis for structural order analysis: (1) quantitative characterization of ring shapes and (2) a spatial correlation function around rings to visualize local structural orders. The advantage of our approach is that it enables a direct combination of the size, shape, and spatial distribution of rings based on both atomic configuration and network topology. The direct approach is effective for deeply understanding intermediate-range structural orders because crucial topological structures formed by chemical bonds are embedded in the atomic configuration. Our methods are applied to analyze amorphous and crystalline silica for a deeper understanding of the structural orders around rings.

**Results**

To reveal the contribution of the network topological order to the intermediate-range structural order, we propose two analysis methods based on rings: (1) ring shape characterization (**Fig. 1**) and (2) spatial correlation analysis around rings (**Fig. 2**).

**Ring shape characterization**

The proposed procedure characterizes the shape of a ring through computation of the eigenvector and eigenvalue of the covariance matrix of atomic coordinates (so-called point cloud) in the ring, as shown in **Fig. 1**. This procedure first selects a ring enumerated from a network consisting of atoms and chemical bonds (**Fig. 1a**) and then obtains a point cloud composed of atoms in the ring (**Fig. 1b**). Subsequently, the eigenvectors and eigenvalues of the variance-covariance matrix of the coordinates are computed (**Fig. 1c**). The first eigenvector is computed as the direction with the largest variance of the point cloud. Under the restriction that the first and second eigenvectors are orthogonal with each other, the second eigenvector is computed by maximizing the variance of the point cloud in the vector. Furthermore, the third eigenvector is computed as the vector orthogonal to the first and second eigenvectors. As their eigenvalues are proportional to the variances of the point cloud along these vectors, they can be used to measure the ring shape. This approach uses eigenvectors, which provide second-order information. This is equivalent to approximating a ring by an ellipsoid, wherein the first and second eigenvectors correspond to the directions of the major and minor axes of the ellipse approximating the ring, respectively, whereas the third eigenvector is the normal vector of the plane of the ellipse. We proposed two shape indicators for a ring: roundness and roughness, as illustrated in **Fig. 1d**, both of which are computed from eigenvalues. Assuming a square root of the three eigenvalues, $s_1$, $s_2$, and $s_3$, the first indicator "roundness" is defined as

$$r_c = \frac{s_2}{s_1}. \tag{1}$$

It evaluates how close the ring is to a perfect circle. The roundness value for a perfect circle is 1. The second indicator "roughness" is defined as

$$r_t = \frac{s_3}{\sqrt{s_1 s_2}}. \tag{2}$$

It evaluates the flatness of the ring. These normalizations, i.e., $s_1$ for roundness and the geometric mean of $s_1$ and $s_2$ for the roughness, are included to remove the effect of the ring size. These indicators can be generalized by weighting atoms in computing ring centers and variance-covariance matrices. Variances of atomic positions weighted by atomic masses, called the radius of gyration, was proposed as a measure to evaluate ring compactions[29].

Examples in **Fig. 1e** demonstrate that roundness and roughness appropriately evaluate these ring shapes. In general, there is no direct relationship between roundness and roughness. For example, roundness and roughness are not correlated for rings with sizes 9 and 11, as shown in **Fig. 1e**. However, they are negatively correlated for rings with sizes 6 and 8. Notably, these numerical values of small rings are restricted due to the small degrees of freedom in their shapes.

**Spatial correlation function around rings**

The second proposed procedure (**Fig. 2**) is used to visualize how the symmetry and/or anisotropy of a ring significantly contribute to the local structural orderings in the intermediate-range scale. For the computational procedure of the spatial correlation function, we first enumerate all the rings (**Fig. 2a**). For each ring, we compute the ring center by averaging atomic coordinates and the eigenvectors of the variance-covariance matrix from the atomic coordinates of the ring (**Fig. 2b**). Then, we obtain the center $\boldsymbol{c}_r = (c_{rx}, c_{ry}, c_{rz})$, and eigenvectors $\boldsymbol{e}_{r1}, \boldsymbol{e}_{r2}, \boldsymbol{e}_{r3}$ for the $r$-th ring, $r = 1, \ldots, N_{\text{ring}}$. Here, $N_{\text{ring}}$ is the number of enumerated rings, while $N_{\text{atom}}$ is the number of an element in the structural model, i.e., the number of Si or O atoms in silica. Next, for each element, i.e. Si or O atom in silica, we compute the spatial three-dimensional (3D) histogram around each ring by three steps (**Fig. 2c**): (1) Atomic coordinates of the element in the structural models $\boldsymbol{x}_n = (x_n, y_n, z_n)^T, n = 1, \ldots, N_{\text{atom}}$ are translated so that the ring center is the origin, which is computed by

$$\bar{\boldsymbol{x}}_{rn} = (x_n - c_{rx}, y_n - c_{ry}, z_n - c_{rz})^T, n = 1, \ldots, N_{\text{atom}} \tag{3}$$

where $\cdot^T$ is a transpose operator of a vector or a matrix. (2) When the coordinate axes of the histogram around $r$-th ring are defined as $\boldsymbol{e}_{r1}, \boldsymbol{e}_{r2}, \boldsymbol{e}_{r3}$, the atomic coordinates in the new coordination system are computed by

$$\boldsymbol{v}_{rn} = (v_{rn1}, v_{rn2}, v_{rn3})^T = (\bar{\boldsymbol{x}}_{rn}^T \boldsymbol{e}_{r1}, \bar{\boldsymbol{x}}_{rn}^T \boldsymbol{e}_{r2}, \bar{\boldsymbol{x}}_{rn}^T \boldsymbol{e}_{r3})^T, n = 1, \ldots, N_{\text{atom}}. \tag{4}$$

(3) The 3D histogram of the element around the $r$-th ring in the new coordination system, $h_{\text{atom},r}(v_1, v_2, v_3)$ is computed using atomic coordinates $\boldsymbol{v}_{rn}, n = 1, \ldots, N_{\text{atom}}$. After computing histograms for all rings, the spatial correlation function of each element, $h_{\text{atom}}(v_1, v_2, v_3)$ is computed by averaging histograms (**Fig. 2d**):

$$h_{\text{atom}}(v_1, v_2, v_3) = \frac{1}{C_{\text{atom}}} \sum_{r=1}^{N_{\text{ring}}} h_{\text{atom},r}(v_1, v_2, v_3). \tag{5}$$

The normalization coefficient is $C_{\text{atom}} = d_w^3 N_{\text{ring}} N_{\text{atom}} V_{\text{box}}^{-1}$, where $d_w$ is the bin width of the histogram, and $V_{\text{box}}$ is the volume of the simulation box. Normalization is necessary to compare structural models with different atomic densities. In crystals, regions with large spatial correlation are point-like and are extremely small for visualization. To improve the visibility of the large correlation regions of crystalline materials, we used a Gaussian filter for their computed correlation functions to spatially expand them.

Our proposed correlation function is a generalized formulation of the results from a recent study[30]. In the study, the spatial correlation function was computed based on a structural unit consisting of three atoms (i.e., a centered Si atom and two nearest O atoms in silica) to determine the coordination system of the correlation function. In contrast, our proposed approach determines the coordination system by using a ring as a larger unit consisting of over four atoms. This offers a more straightforward approach for visualizing and revealing intermediate-range structural orders in a network-forming amorphous material, which is demonstrated by identifying similarities in structural orders between amorphous and crystalline materials of silica.

**Analysis of amorphous and crystalline silica structural models**

In this study, we focused on amorphous and crystalline materials with corner-sharing tetrahedral motifs, which are represented by a structural model of amorphous silica ($a$-SiO$_2$), along with those of nine crystalline polymorphs of silica: $\alpha$-tridymite[31], $\beta$-tridymite[32], $\alpha$-cristobalite[33], $\beta$-cristobalite[34], $\alpha$-quartz[35], $\beta$-quartz[36], coesite I[37], coesite II[38], and stishovite[39] (see **Supplementary Tables 1–3** for data IDs, space groups, and lattice information, and **Supplementary Discussion** and **Supplementary Figure 1** for structural statistics). Note that the coordination number of Si is six only in stishovite, while it is four in other crystals analyzed in this work.

We generated a large-scale structural model of $a$-SiO$_2$ by classical molecular dynamics (MD) simulation of a melt-quenching procedure, followed by refinement using the reverse Monte Carlo (RMC)[40] technique to reproduce XRD and ND data. Here, the side length of the simulation box for the model was assumed approximately 100 Å to ensure accurate statistics and suppress artifacts caused by the periodic boundary condition (see **Methods** and **Supplementary Table 4** for the detailed procedure and structural statistics). RMC was implemented with some constraints to preserve the physically meaningful structure generated by MD (**Methods** and **Supplementary Discussion**). Hereafter, the generated structure is referred to as the MD–RMC model. The X-ray and neutron total structure factors $S(q)$ of the model showed good agreement with the experimental data (**Supplementary Figures 2** and **3**). The coordination numbers around Si and O atoms, of over 99% in the first coordination distance of the generated model, are four and two, respectively. It was established that a well-known network structure, consisting of an SiO$_4$ tetrahedron sharing O atoms at the corner for SiO$_2$ ($N_{\text{Si}-\text{O}} = 4$, $N_{\text{O}-\text{Si}} = 2$), is formed in the MD–RMC model (**Supplementary Figures 4** and **5)**.

**Ring shape characterizations**

Guttman ring[15], King ring[17], and primitive ring[18,19] are well-known ring definitions. We selected the primitive ring for our investigations. A primitive ring is defined as a ring that cannot be decomposed into two smaller rings[18,19]. This is equivalent to the condition that for any node pairs in the ring, there is no shortcut, which is a shorter path composed of edges outside the ring. Therefore, primitive rings are essential components in network-forming amorphous materials. Although the primitive criterion enumerates larger rings compared to Guttman's rings, it can avoid enumerating redundant larger rings unlike those in King's criteria. Please see Yuan's paper[19] for some illustrative examples of primitive rings. As a first step, our analysis exhaustively enumerated rings in each network of the structural models based on the shortest-path algorithms (See **Methods** for the detail.). **Fig. 3a** shows the distributions of ring sizes, i.e., the numbers of Si atoms in a ring in amorphous and crystalline materials. Among the crystalline materials, cristobalite and tridymite phases have only six-fold rings, although the coordination number of each Si and O atom is the same in all crystalline materials. Therefore, these structures are topologically ordered even while considering the network connectivity patterns over the first coordination spheres. In contrast, other crystal phases have rings of different sizes. Notably, the ring sizes of coesite I and II are widely distributed, as in the amorphous material shown in the bottom panel, thus indicating that the structures of coesite I and II are topologically disordered, owing to its high density. The ring size of $a$-SiO$_2$ is also widely distributed, with the peak of the ring size distribution located at approximately sizes 6 or 7, similar to the cases of phases of cristobalite and tridymite.

For these structural models, we applied the proposed ring-shape characterizations, as summarized in **Fig. 1**. **Fig. 3b–c** show our proposed indicators of ring shape, i.e., roundness and roughness, in crystalline and amorphous silica materials. It was found that $\beta$-cristobalite has only six-fold rings with the largest roundness and smallest roughness compared to other crystals. Thus, compared to other crystalline materials, $\beta$-cristobalite has the highest ring symmetry, which contributes to the cubic nature of the crystal system, as discussed later. In contrast, the rings of $\alpha$-cristobalite are less symmetric than those of $\beta$-cristobalite but with higher symmetry than those of coesite I and II. Among these structures, the order of the ring symmetries is identical to the order of the ring symmetries in crystal systems.

We demonstrated that a set of three eigenvalues of a ring can be used as an indicator of the ring's shape. Notably, the set is also useful for enumerating isomorphic rings in crystalline materials because these values are identical when the ring shapes are identical. The isomorphic rings identified in crystalline SiO$_2$ and their shape characterizations are summarized in **Fig. 4** and **Supplementary Table 5**. Both $\beta$- and $\alpha$-cristobalite have only one isomorphic ring, whereas $\beta$- and $\alpha$-tridymite, $\beta$- and $\alpha$-quartz, coesite I, coesite II, and stishovite have two, three, four, five, eleven, thirty-four, and four isomorphic rings, respectively. This indicates that the number of isomorphic rings in a crystalline material is a measure of its symmetry.

Next, we analyzed linkages between ring symmetries and space groups of crystals. As a factor of the ring symmetry, the point group of each isomorphic ring is listed in **Supplementary Table 5**. The result concludes that only $\beta$-cristobalite and $\beta$-tridymite have a direct linkage between the point group of a ring and the space group of a crystal. For example, $\beta$-cristobalite includes only an isomorphic ring (A6-1) with the highest symmetry evaluated by both point group and our shape measures, as shown in **Supplementary Table 5**. We found that the three-fold rotational symmetry of the ring in $\beta$-cristobalite directly links to the three-fold symmetry of the cubic space group, as illustrated in **Fig. 5a–b**. Because $\beta$-cristobalite includes only one isomorphic ring, this situation leads to the formation of a cubic crystal system that has the highest symmetry among the crystalline materials of $SiO_2$. We also found a direct linkage in $\beta$-tridymite, as illustrated in **Fig. 5c–e**. The ring with three-fold symmetry (C6-1), whose shape is identical to that of the ring (A6-1) in $\beta$-cristobalite, can yield six-fold screw symmetry combined with the partial translational operation along the screw axis. Lastly, rings in $\alpha$-cristobalite, $\alpha$-tridymite, $\beta$-quartz, $\alpha$-quartz, coesite I, and coesite II are unrelated to the space groups. Although $\alpha$-cristobalite includes only an isomorphic ring similar to $\beta$-cristobalite, the point group indicates no symmetry, small roundness, and large roughness, resulting in a less symmetric space group than that of $\beta$-cristobalite, as illustrated in **Supplementary Figure 6**. In this case, the configuration of rings with the lowest symmetry constructs four-fold screw symmetry. A similar situation is also found in $\alpha$-tridymite, $\beta$-quartz, and $\alpha$-quartz, as illustrated in **Supplementary Figures 7–9**. The cases of coesite I and coesite II become extremely complicated, and it is hard to find any linkages between ring symmetries and space groups in such cases. Overall, this analysis concludes that there is a strong linkage between the ring symmetry and the crystal symmetry, i.e., highly symmetric rings build a highly symmetric crystal system.

We further investigated the linkage between ring symmetries and crystal polymorphs based on the pressure–temperature phase diagram, as shown in Fig. 4 in a previous paper[41]. Two noteworthy findings regarding the phase diagrams are as follows: first, that high-temperature phases are more symmetric than the low-temperature ones. The symmetry is evaluated by the number of isomorphic rings and ring symmetries (such as roundness, roughness, and point group). For example, $\beta$-phases (such as $\beta$-cristobalite) are more symmetric than $\alpha$-phases (such as $\alpha$-cristobalite). In addition, cristobalites are more symmetric compared with tridymites. Second, except for stishovite, which has different atomic coordination numbers from other crystals, low-pressure phases are more symmetric than the high-pressure ones. For example, phases of cristobalite and tridymite are more symmetric than those of quartz and coesite. These two findings are consistent with the fact that $\beta$-cristobalite is the most symmetric among $SiO_2$ crystal polymorphs since the phase is located at the lowest pressure and highest temperature in the phase diagram.

The bottom panels in **Fig. 3b**–**c** show the computed results for $a$-$SiO_2$. Similar to the ring size distribution, roundness and roughness are broadly distributed. This indicates that various rings are included in $a$-$SiO_2$. The ring-shape evaluation with two-dimensional distributions of both measures is shown in **Fig. 6**. The probability density in $a$-$SiO_2$ was computed by a kernel density estimation with Gaussian kernel, whose band width was determined using the Scott's rule[42]. Characterization results for crystalline materials are represented by symbols due to the limited number of isomorphic rings. **Fig. 6** shows that these values of $\alpha$-cristobalite, $\beta$-cristobalite, $\alpha$-tridymite, and $\beta$-tridymite are negatively correlated because of the small ring sizes. However, a correlation is not generally observed, instead, it is not adaptable for most cases. The figure shows that the peak position of the roundness–roughness distributions in $a$-SiO$_2$ is close to those of the ring B6-1 in $\alpha$-cristobalite and D6-3 in $\alpha$-tridymite among the six-fold rings. This shows a resemblance between the major rings in $a$-SiO$_2$ and those in $\alpha$-cristobalite and $\alpha$-tridymite. The broad distribution of the shape indicators in $a$-SiO$_2$, covering those of all isomorphic rings in crystalline materials, suggests that the distribution in $a$-SiO$_2$ is a mixture of isomorphic rings in crystalline materials and their variants.

**Spatial correlation functions around rings**

The influence of ring symmetries on the local structural ordering was evaluated using our proposed spatial correlation function. To visualize the functions of SiO$_2$ materials, corresponding cross-sectional mappings at approximately $v_1 = 0$, $v_2 = 0$, or $v_3 = 0$ were computed by integrating the correlation function over the cross-section thickness $t$, as shown in **Fig. 7a–j**. The cross-sectional thicknesses $t$ was set to 2 Å for all crystalline and amorphous models. In these figures, the blue and red regions exhibit high densities of Si and O atoms, respectively, as depicted by the color indicator (**Fig. 7k**). These mappings also show that $\beta$-, $\alpha$-cristobalite, $\beta$-tridymite, and $\alpha$-tridymite (**Fig. 7a–d**) have parallel planes along the $v_3$-axis and concentric circles on the $v_1$-$v_2$ plane, whereas other crystalline materials (**Fig. 7e–i**) appear to be disordered. This is attributed to the fact that phases of cristobalite and tridymite have a smaller number of isomorphic rings with relatively larger roundness and smaller roughness than other crystalline materials. Among all crystalline materials, the function of $\beta$-cristobalite (**Fig. 7a**) is highly oriented and are surprisingly identical due to only one isomorphic ring (six-fold ring) with large roundness and small roughness. In contrast, the functions for $\alpha$-cristobalite, $\beta$-tridymite, and $\alpha$-tridymite (**Fig. 7b–d**) appear noisy compared to that of $\beta$-cristobalite. This is because the three crystalline materials have a larger number of isomorphic rings or smaller roundness and larger roughness of rings than those of β-cristobalite. Coesite II (**Fig. 7h**) is the most disordered due to the variety of non-symmetric isomorphic rings. From another visualization of 3D surfaces of large spatial correlations shown in **Supplementary Figure 10**, we could have similar findings and discussions.

    **Fig. 7j** shows the spatial correlation function for the $a$-SiO$_2$ model. These diagrams exhibit a ring around the center, which is the average of all primitive rings. The $v_1$-$v_2$ plane also shows that the averaged ring shape in $a$-SiO$_2$ is not a perfect circle, but an ellipse, which is consistent with the results of the roundness analysis shown in **Fig. 3b**. As shown in **Fig. 7j** and **Supplementary Figure 10j**, the center ring is surrounded by multiple shells in the amorphous model. The shells of the O and Si atoms are alternately located from the center owing to the O–Si–O–Si linkage. In addition, these figures show planes parallel to the center ring in the upper and lower parts of the ring along the $v_3$-axis. They also show partially parallel planes along the $v_2$-axis, which is parallel to the $v_1$-axis (the major axis of the rings), as shown in the middle panel of **Fig. 7j**. However, they do not exhibit such planes along the $v_1$-axis. Owing to the anisotropic shapes of the rings, it can be concluded that $a$-SiO$_2$ has anisotropic structural order around the rings. In particular, the planes consisting of rings are parallel to each other but orthogonal to the normal vector of a ring. Since this situation is similar to the arrangement of lattice planes in crystals such as β-cristobalite and β-tridymite, as shown in **Fig. 5**, the pseudo-Bragg condition for diffraction could also be considered locally even in amorphous materials. In fact, the normal vector of the ring becomes parallel to the scattering vector. Since the spacing between the planes (~ 4Å) corresponds to the position of FSDP in reciprocal space, the origin of FSDP is likely the formation of intermediate-range structural orders containing these parallel rings.

We further investigated the consistency between the spatial correlation functions of the amorphous and crystalline models. **Supplementary Figure 11** shows the cross-sectional mappings of the Si correlation functions in the amorphous model together with those in crystalline models: $\beta$-cristobalite (**a**), $\alpha$-cristobalite (**b**), $\beta$-tridymite (**c**), $\alpha$-tridymite (**d**), $\beta$-quartz (**e**), $\alpha$-quartz (**f**), coesite I (**g**), coesite II (**h**), and stishovite (**i**). In these diagrams, the function of the amorphous model is shown by the blue image, whereas that of the crystal model is shown by a green image. **Supplementary Figure 12** shows correlation functions of the O atom, wherein spatial correlations of the amorphous model and those of crystalline models are colored red and green, respectively. The rightmost diagrams ($v_1$-$v_2$ plane) show that the averaged ring of the amorphous model is similar to that of $\alpha$-cristobalite (**Supplementary Figures 11b and 12b**), which is consistent with the roundness of rings in **Fig. 3b**. The averaged ring shape of $\alpha$-tridymite (**Supplementary Figures 11d and 12d**) is similar to that of a-SiO$_2$, whereas those of $\beta$-cristobalite (**Supplementary Figures 11a and 12a**) and $\beta$-tridymite (**Supplementary Figures 11c and 12c**) are different, but both are consistent with these roundness values. The upper and lower planes made by the large correlation regions for $a$-SiO$_2$, which are pseudo-Bragg planes, are also consistent with those for the phases of cristobalite and tridymite. The correlation functions for the phases of quartz and coesite (**Supplementary Figures 11e–h and 12e–h**) also contain regions partly similar to those for $a$-SiO$_2$, although they appear different globally. The anisotropic nature of the $a$-SiO$_2$ amorphous model is consistent with the analysis results for crystalline SiO$_2$, especially with phases of cristobalite and tridymite.

For the amorphous SiO$_2$, we performed an additional analysis to find a linkage between the roundness and roughness of rings and the symmetry of the intermediate-range atomic configuration around rings. To confirm the linkage, the $\beta$-cristobalite, which has the highest crystal symmetry and contains one type of symmetric ring with large roundness and small roughness, is used as a standard. In the analysis, we first selected four subsets of rings using their roundness and roughness values, as shown in **Fig. 8a**. The subsets (or regions in the figure) are determined by the quantiles of roundness $r_c = 0.73, 0.80, 0.87$, and those of roughness $r_t = 0.24, 0.30, 0.36$. For example, the range of region I is $0.87 < r_c \leq 1$ and $0 \leq r_t \leq 0.24$. Rings in region I are more symmetric than those in other three regions because of larger roundness and smaller roughness of rings in the region. On the contrary, rings in region IV are less symmetric than others. For each region, we computed the spatial correlation function using rings included in the region. **Fig. 8b** shows the computed spatial correlation functions, along with the correlation function of $\beta$-cristobalite to discuss a symmetry of the intermediate-range atomic configurations around the rings. Direct comparisons of these functions of amorphous $SiO_2$ with that of $\beta$-cristobalite in same diagrams are shown in **Supplementary Figure 13b**. These figures demonstrate that $\beta$-cristobalite and the functions of symmetric rings similar to a perfect circle (the spatial correlation function for region I in **Fig. 8b**) and non-symmetric rings that are dissimilar to a perfect circle (the function for region IV in **Fig. 8b**) in $a$-$SiO_2$. This demonstrates that atomic configurations are locally symmetric around symmetric rings, even in $a$-$SiO_2$ and that the lesser the symmetric rings, the lesser symmetric the local atomic configurations, as shown in spatial correlation functions for regions II–IV in **Fig. 8b** and **Supplementary Figure 13b**. In $a$-$SiO_2$, highly symmetric rings are scarce, while slightly less symmetric rings are predominant in region II, which includes the probability density mode. Therefore, the amorphous material is mainly composed of slightly less symmetric local structures in the present case.

**Comparing amorphous models with various structural orderings**

A series of analyses showed an anisotropic local structural order in the amorphous silica (*a*-SiO$_2$) model, whereas isotropic halo rings were observed in the reciprocal space by macroscopic diffraction data. A transition from anisotropic to isotropic order prevails while transitioning from microscopic to macroscopic viewpoints. Notably, the anisotropic local structure originates from the rings, which is an essential topological motif in network-forming materials. Here, we discuss the influence of the anisotropic local structural order on the degree of structural order in amorphous structures through the comparison of some additional amorphous models.

To validate our analyses further, we generated three additional amorphous structural models from a random configuration. The same number of atoms and the same size of the simulation box were used for these additional models. The first model (Rand-Coord) was generated from a random configuration, followed by a hard-sphere Monte Carlo (HSMC) simulation under two restrictions: the closest atom–atom distance and coordination number. The closest-distance restriction avoids unreasonable spikes in the partial pair distribution functions. The coordination number restriction forces Si atoms to coordinate to four O atoms, whereas O atoms coordinate to two Si atoms within a Si–O cutoff distance of 1.90 Å. The second structural model (Rand-Tetra) used the Rand-Coord model as the initial atomic configuration and was then generated by implementing HSMC with restrictions on the coordination numbers and O–Si–O bond angle distribution to create a network structure, consisting of regular SiO$_4$ tetrahedra sharing O atoms at the corner. The third model (Rand-RMC) used the Rand-Tetra model as the initial atomic configuration, subsequently generated by the RMC simulation to reproduce the X-ray and neutron $S(q)$ data. Structural statistics, such as structure factors, coordination numbers, bond angles, and ring characters are summarized in **Supplementary Discussion** and **Supplementary Figures 14–17**. This structured summary demonstrates that the randomly initialized models (i.e., Rand-Coord, Rand-Tetra, and Rand-RMC) are more disordered compared to the MD-RMC model. Among the structure factors, only Rand-RMC and MD-RMC exhibit FSDP because RMC fits the structure factors. Although all models form a network of chemical bonds, the structural orders of these amorphous models are highly different. Structures of three models without MD (Rand-Coord, Rand-Tetra and Rand-RMC) are considerably more disordered than MD-RMC models, as demonstrated in **Supplementary Figure 17**, wherein all variances of ring size, roundness, and roughness of these three models are larger than those of the MD-RMC models. These statistics indicate that RMC does not construct structural order, instead, it generates disordered structures[43].

**Fig. 9** shows the spatial correlation functions of Rand-Coord (**a**), Rand-Tetra (**b**), Rand-RMC (**c**), and MD-RMC (**d**), wherein all models exhibit anisotropic structural orders around the rings. This indicates that anisotropy is related only to the network formation of amorphous materials. A clearer shell structure was observed in the MD-RMC model compared to other models. This is due to the ring size distribution in the MD-RMC model, concentrated at size 6 (see **Supplementary Figure 18**), whereas other three models showed broad distributions. Hence, the correlation functions computed using only six-fold rings, as shown in **Supplementary Figure 18**, present clearer shell structures in all models. The correlation functions for Rand-RMC and MD-RMC in **Figs. 9c–d** and **Supplementary Figures 18c–d** exhibit parallel planes above and below rings along the $v_3$-axes indicated by green arrows. This relates to the intermediate-range structural orders exhibited by FSDP in the diffraction experiments. Conclusively, the anisotropy of structural orders around rings is essential for building intermediate-range structural orders.

**Discussion**

This study proposed two analysis methods for the structural orders of rings and those around rings based on the ring shapes in network-forming materials. The proposed analysis approach for crystalline and amorphous $SiO_2$ first demonstrated that the distribution of ring shape characteristics is a strong tool for analyzing the amorphous models of $SiO_2$. The roundness–roughness distribution showed that the major rings in amorphous $SiO_2$ are similar to those in two crystals, $\alpha$-cristobalite and $\alpha$-tridymite, whose mass densities are similar to those of amorphous $SiO_2$. In addition, it was found that the rings in quartz and coesite phases are included as minor rings in amorphous $SiO_2$. This indicates that various rings found in crystalline polymorphs are necessary to form bulk amorphous materials[44]. Furthermore, the spatial correlation analysis showed that the anisotropic nature of structural orders was found in both crystalline $SiO_2$, with different polymorphisms, and amorphous $SiO_2$. The analysis further showed that pseudo-Bragg planes, parallel to the rings in amorphous $SiO_2$, are consistent with those of four crystals: $\alpha$-cristobalite, $\alpha$-tridymite, $\beta$-cristobalite, and $\beta$-tridymite. In addition, this study found that rings with low roughness tend to construct planes parallel to the central rings, as shown by the ring shape characterization of isomorphic rings on crystalline materials. To validate the applicability of our approaches, three amorphous $SiO_2$ structural models with different ordering degrees were prepared. The corresponding anisotropic local environments, as described above, were confirmed in all models. Although the actual degree of order for amorphous $SiO_2$ is still arbitrary, observations in this study showed that even if amorphous $SiO_2$ satisfies the experimental structure factors, it has anisotropic local structures. Furthermore, our analysis assumed that the observed pseudo-Bragg planes, presumably yielding the first sharp diffraction peak, are correlated with the intermediate-range structural orders in amorphous $SiO_2$.

To link structural orders and material properties and realize data-driven material design, quantitative evaluations for various structural models are necessary. In crystalline materials, the macroscopic properties are linked to the lattice structure and their symmetry because they include all structural information. Conversely, amorphous materials do not have such linkages because they do not have the lattice and symmetric structures. Therefore, we should explore the linkages through another approach using structural motifs such as rings. An example of this approach is using different fictive temperatures[45], which revealed the linkage between the properties and ring size distributions in $SiO_2$ glass. Various additional numerical descriptors such as geometric information of rings like roundness and roughness, and those of structural ordering from diverse perspectives are needed for finding new linkages between macroscopic properties and structural orders. For network-forming materials, we developed new descriptors and demonstrated useful applications for analyzing the structural orders of $SiO_2$ materials, although the applicability of the approach is not limited to $SiO_2$. Furthermore, the analysis of other network-forming materials using our proposed methods and a comprehensive comparison of the results should provide further information to help understand intermediate-range structural orders in network-forming materials.

**Methods**

**Generation of a structural model of amorphous silica**

Structural models of amorphous silica were generated by a melt-quenching procedure in classical MD simulation, followed by RMC techniques to fit the diffraction measurements of the synthesized materials[40]. MD simulations and RMC were implemented using the LAMMPS code[46] and the RMC++ code[47], respectively. In the MD simulation of amorphous silica, the simulation box was a cube with a length of 100 Å. The density of this system was 2.2 g cm$^{-3}$. The system had 66,156 atoms (22,052 Si atoms and 44,104 O atoms) within the NVT ensemble. A time step of 1 fs was used in the Verlet algorithm. Furthermore, the interactions were described by pair potentials with short-range Born-Mayer repulsive and long-range Coulomb terms:

$$\phi_{ij} = B_{ij} \exp\left(-\frac{r}{\rho_{ij}}\right) + \frac{e^2}{4\pi\epsilon_0} \frac{Z_i Z_j}{r}, \tag{6}$$

where $r$ is the interatomic distance between the atoms $i$ and $j$, $B_{ij}$ and $\rho_{ij}$ represent the magnitude ($10^{-16}$ J) and softness (Å) of the Born-Mayer term, respectively, $Z_i$ is the effective charge on atom $i$ ($Z_{Si} = 2.4$ and $Z_O = -1.2$), $e$ is the elementary charge, and $\epsilon_0$ is the permittivity of a vacuum. The coefficients used in the simulation were $B_{ij} = 21.39 \times 10^{-16}$ J and $\rho_{ij} = 0.174$ Å for an atomic pair of Si and O, and $B_{ij} = 0.6246 \times 10^{-16}$ J and $\rho_{ij} = 0.362$ Å for a pair of O atoms. In the potential function, the interactions between the Si atoms were ignored, similar to our previous work[20]. Although numerous studies have investigated the empirical potential functions for $SiO_2$ materials, potentials that assume Si–Si interaction are limited. One example is the reference study[48], wherein the potential function with Si–Si interactions was determined using ab Initio quantum-chemical methods. However, the effect of Si–Si interactions is smaller than those of other interaction terms. A comparison of these potentials reported in another study[49] showed that the potential function without Si–Si interactions is more accurate than that with Si–Si interaction. Because of the discussion in previous works, we considered an empirical potential function without Si–Si interaction.

The atomic configuration was initialized at random, and the system was equilibrated at 4,000 K for 100,000 steps. Subsequently, it was cooled to 300 K for 5,000,000 steps and annealed at 300 K for 100,000 steps. The generated model was refined via RMC modeling using XRD and ND measurements. RMC was implemented with the constraints of the coordination numbers, bond angles of O–Si–O, and the partial pair-distribution functions within the first coordination shell. The constraint of the bond angle was performed by preserving the distribution of the structures by the MD simulation. These constraints were added to preserve the physically meaningful structure generated by MD, which is discussed in detail along with some structural statistics. This model was named MD-RMC.

**Ring enumeration**

Under the assumption that the first coordination distance is less than 2.0 Å for Si–O, a network was generated for each structural model of $SiO_2$. Primitive rings can be efficiently enumerated based on the shortest-path algorithms[19]. The enumeration algorithm first computes the distances of the node pairs in the network using the shortest path algorithm. Next, node pairs whose distances in the network are less than the threshold, which should be set to half of the maximum number of atoms counting both Si and O atoms in the primitive rings, are enumerated. Subsequently, all the shortest paths between the enumerated node pairs are enumerated. Ring candidates are generated by connecting two different shortest paths that do not share internal nodes. Upon inspecting the primitive criterion for the generated candidates and removing candidates that do not satisfy the criterion, the ring enumeration algorithm is terminated. The ring enumerations in this study were implemented using our in-house python package.

**Data availability**

All data needed to evaluate the conclusions in the paper are present in the paper and/or Supplementary information. Additional data related to this paper may be requested from the authors.

**Code availability**

The codes developed during the current study are available from the corresponding author on reasonable request.

**Acknowledgments**

This research was supported by JSPS KAKENHI Grant Numbers JP20H05878 (to M.S.), JP20H05884 (to M.S.), JP23K17837 (to A.H.), JP20H05881 (to A.H. and Y.O.), JP20H05882 (to H.M.), JP20H04241 (to M.S., A.H., and Y.O.), and JP19K05648 (to Y.O.) and JST PRESTO Grant Number JPMJPR16N6 (to M.S.). We would also like to thank Dr. Shinji Kohara for the helpful discussions.

**Author contributions**

M.S., A.H., and Y.O. designed the study. M.S. developed the analysis methods and computational codes. M.S. and Y.O. built the structural models and analyzed the configurations. A.H. designed the analysis on the linkage between ring symmetries and structural symmetries, and A.H. and M.S. performed them. M.S. prepared a draft of the manuscript. M.S., A.H., Y.O., and H.M contributed to the writing and editing of this manuscript.

**Competing interests**

The authors declare no competing interests.

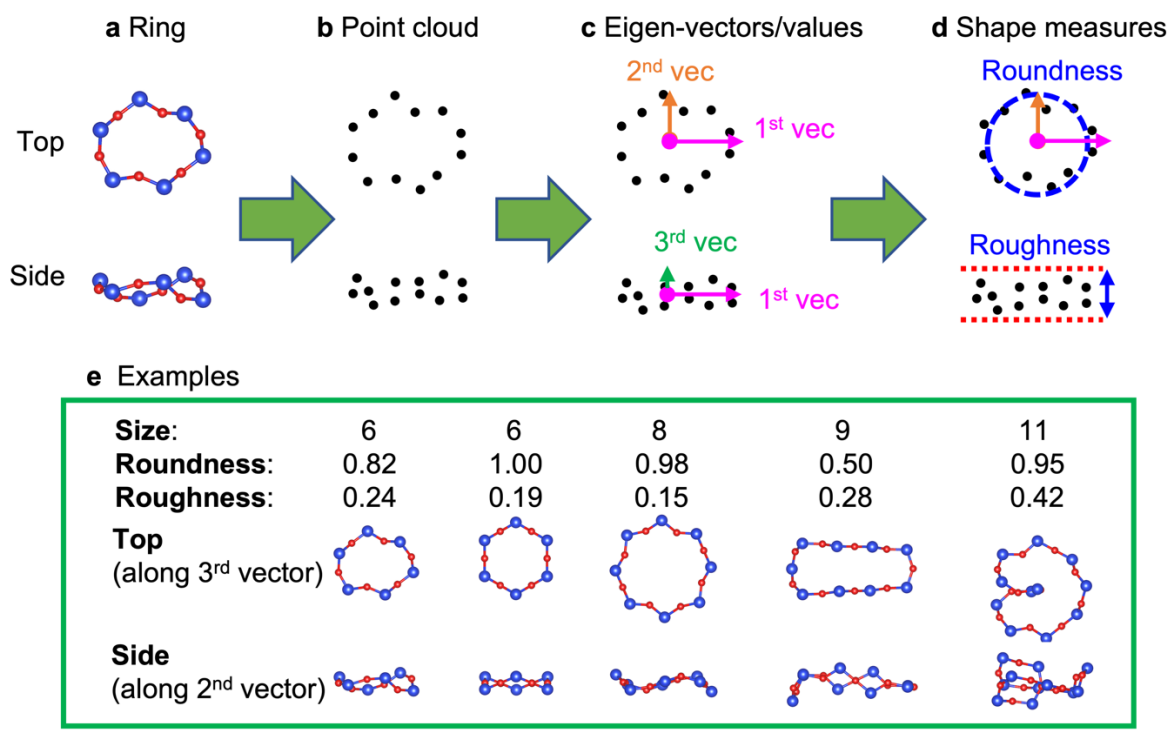

**Fig. 1. Computational procedure of ring shape characterizations.**
 **a** Rings observed from top and side views. **b** The point cloud of atoms in the ring. **c** The first, second, and third eigenvectors and eigenvalues of the variance-covariance matrix of the point cloud. **d** Ring shape measurements: roundness and roughness, both of which are computed from the eigenvalues. Roundness and roughness are defined as Eqs. (1) and (2), respectively. **e** Examples of ring characterizations by ring size, roundness, and roughness for various rings along with observation from the top (the third eigenvector) and that from the side (the second eigenvector).

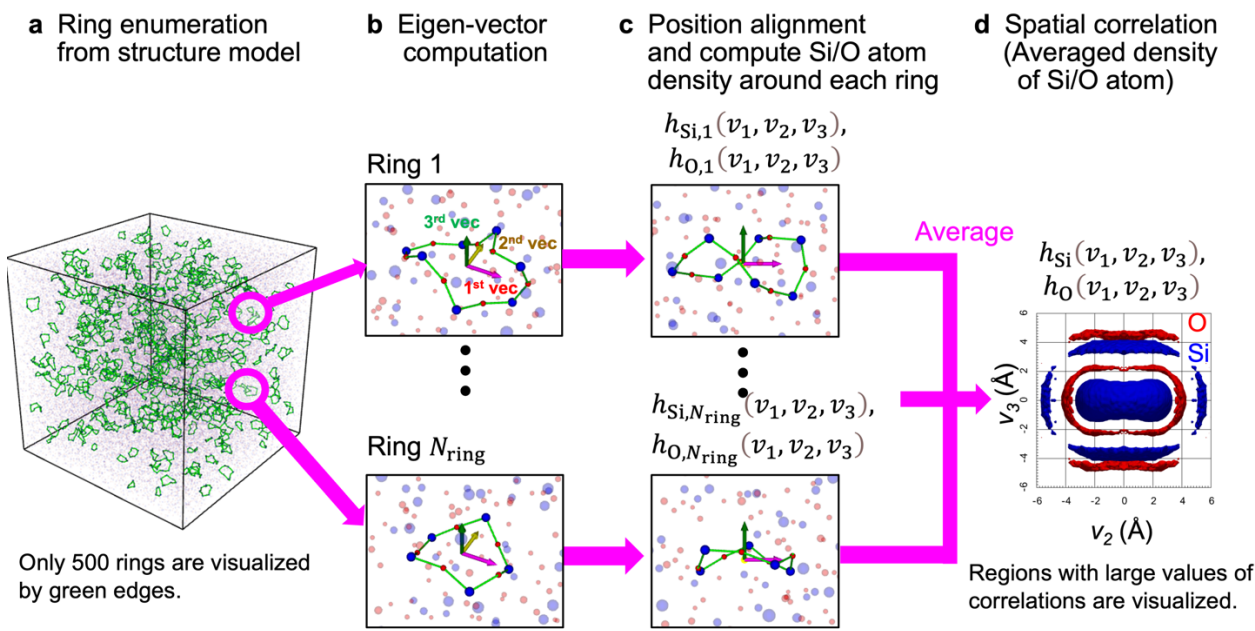

**Fig. 2. Computational procedure of spatial correlation function around rings.**
**a** Illustration of a network topology of amorphous $SiO_2$. Blue and red spheres represent Si and O atoms, respectively. Only 500 enumerated rings chosen from $N_{ring}(= 45,423)$ rings at random are visualized by highlighting their edges in green. **b** Eigenvectors of the variance-covariance matrix from atomic coordinates in each ring are computed. **c** For each ring, the new coordination system $(v_1, v_2, v_3)$ is defined by eigenvectors. The value of $v_d$ indicates the relative position from the ring center along the $d$-th eigenvector. Atomic coordinates in the system are computed by Eqs. (3) and (4). Next, for each ring $r$, spatial histograms for Si or O atom $h_{Si,r}(v_1, v_2, v_3)$, $h_{O,r}(v_1, v_2, v_3)$ are computed using atomic coordinates in the new system. **d** Spatial correlation functions of Si or O atoms around rings $h_{Si}(v_1, v_2, v_3)$, $h_O(v_1, v_2, v_3)$ are computed by averaging the spatial histograms, as expressed in Eq. (5).

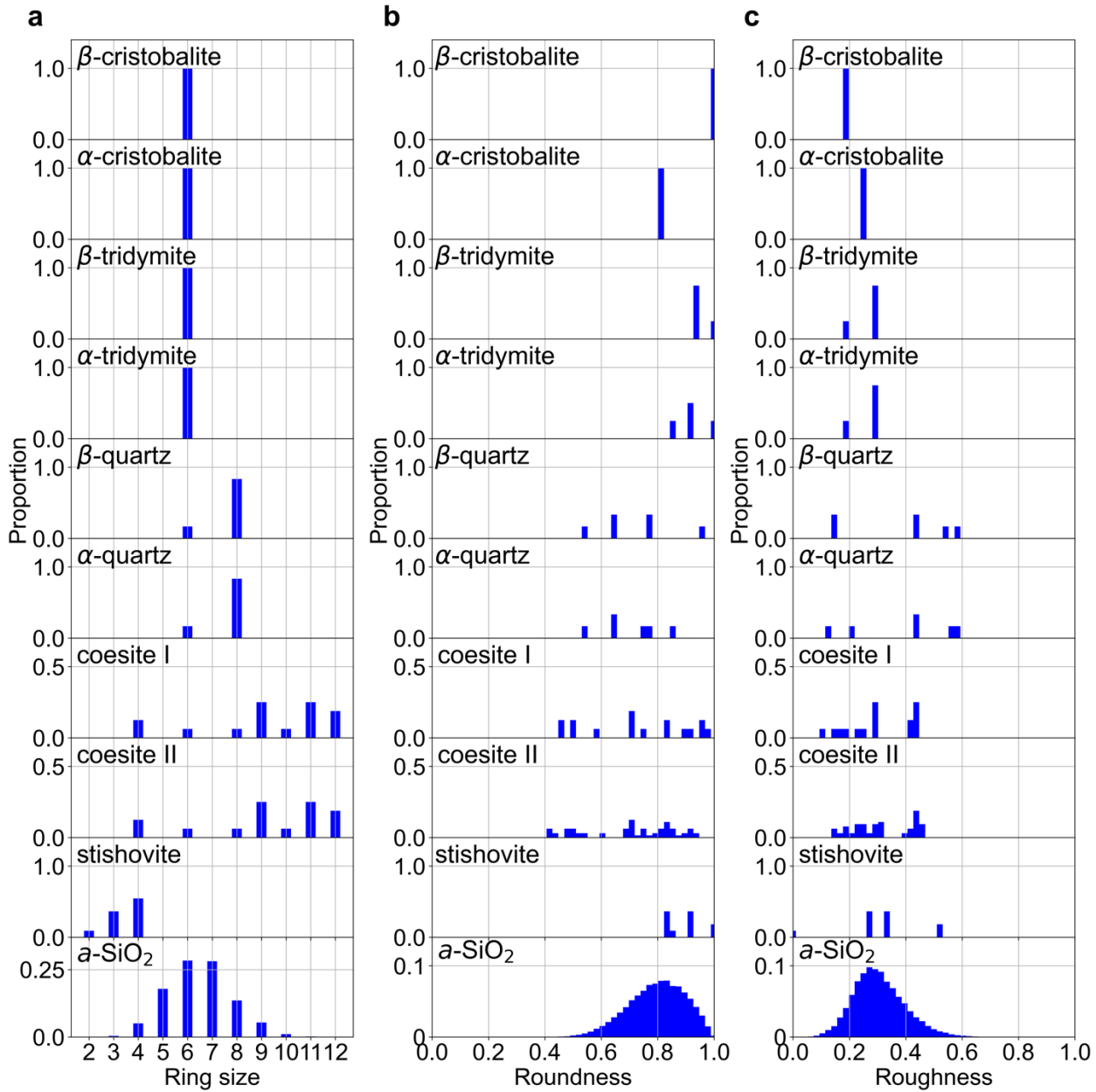

**Fig. 3. Ring characterizations of SiO$_2$ materials.**

Proportion of the number of rings in crystalline and amorphous materials as a function of ring size (**a**), roundness (**b**), and roughness (**c**). Ring size is defined by the number of Si atoms in a ring. Roundness and roughness of a ring are defined as Eqs. (1) and (2), respectively.

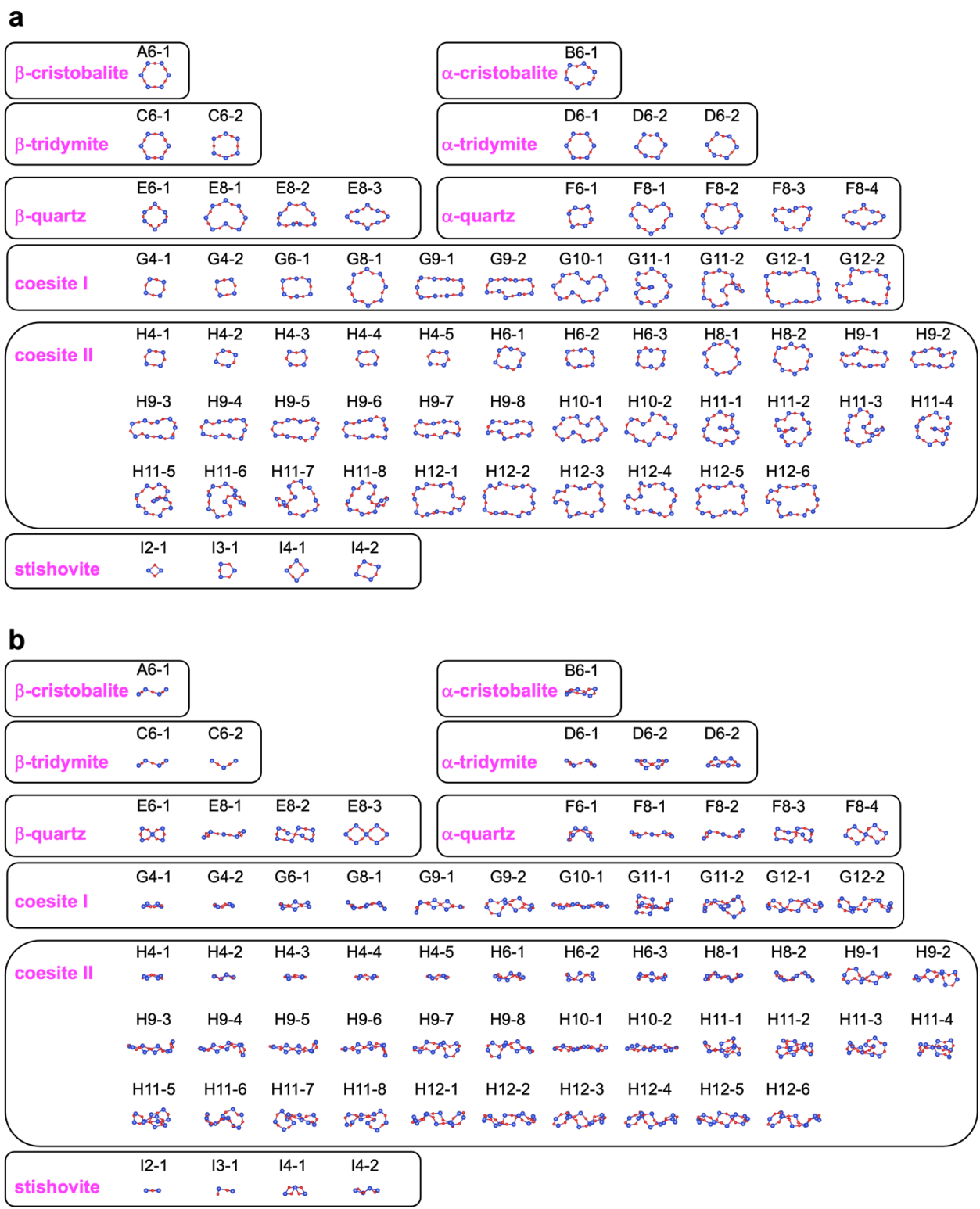

**Fig. 4. Isomorphic primitive rings in crystalline SiO$_2$.**
  **a** The top view (the normal vector) of rings. **b** The side view (the second eigenvector). Numerical values of ring shape characteristics and the point group of each ring are listed in **Supplementary Table 5**.

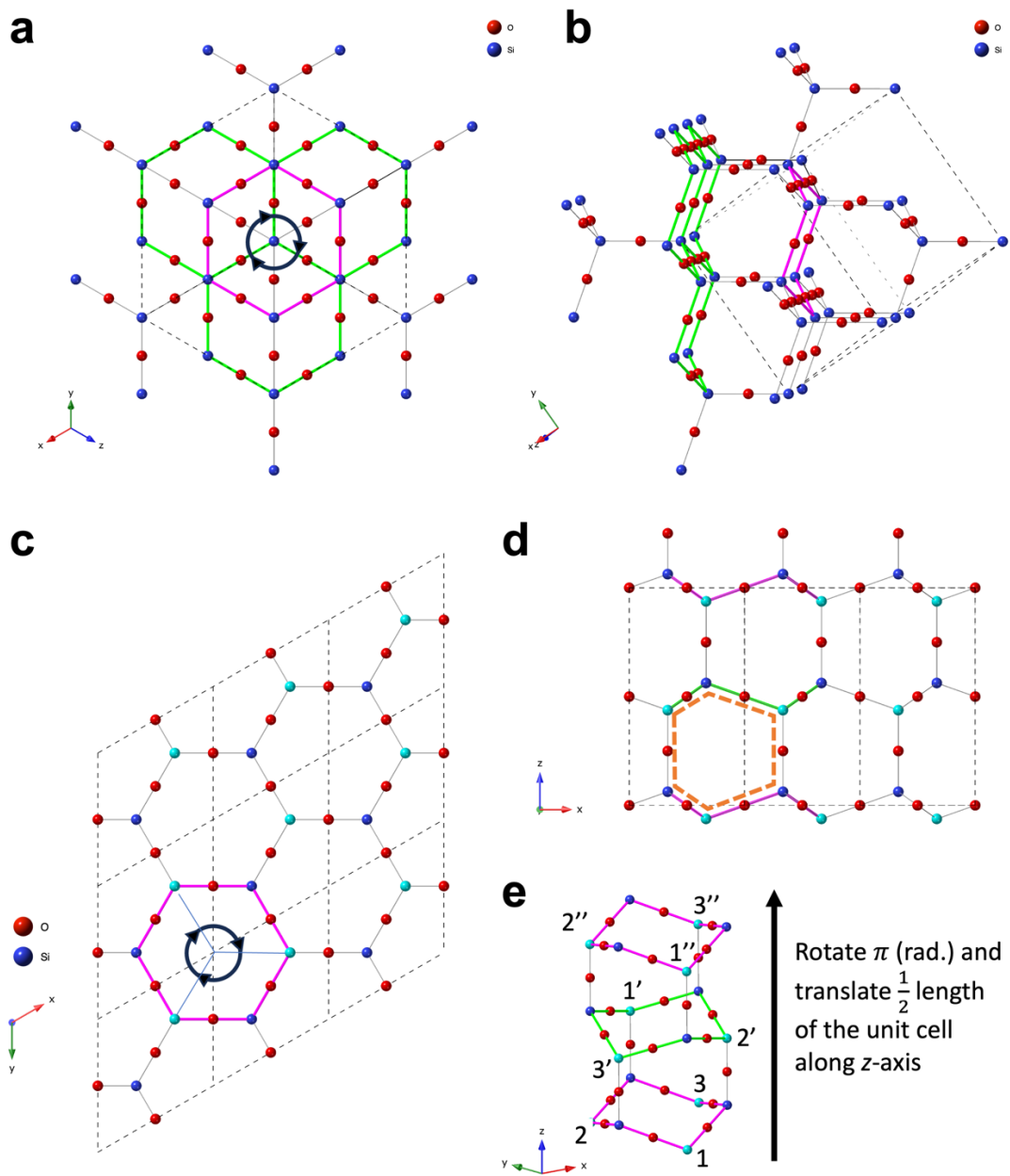

Fig. 5. **Symmetry with rings in *β*-cristobalite and *β*-tridymite. a** The view from the direction [111]. **b** The side view of that in (**a**). In these panels, all rings are identical to the ring A6-1 in **Supplementary Table 5**, some of which are colored magenta or green to visualize layers. This visualization shows the direct linkage between point group of the ring ($\bar{3}$ m) to the space group of *β*-cristobalite (F d $\bar{3}$ m). **c** The view from the direction [001] of *β*-tridymite. The magenta ring is C6-1, whose point group is $\bar{3}$ m. **d** The view from the direction [010]. The orange ring is C6-2. All rings of C6-2 are orthogonal to ring C6-1. Si atoms are colored blue or cyan, which indicate site equivalence whether the *z* coordinate is lower or higher from the ring plane. **e** Screw symmetry $6_3$ in *β*-tridymite, which is along the direction [001] with C6-1 rings. Numbers marked in prime and double prime indicate trajectories of the screw operation, which demonstrates the direct linkage between the point group of the ring and the space group of *β*-tridymite.

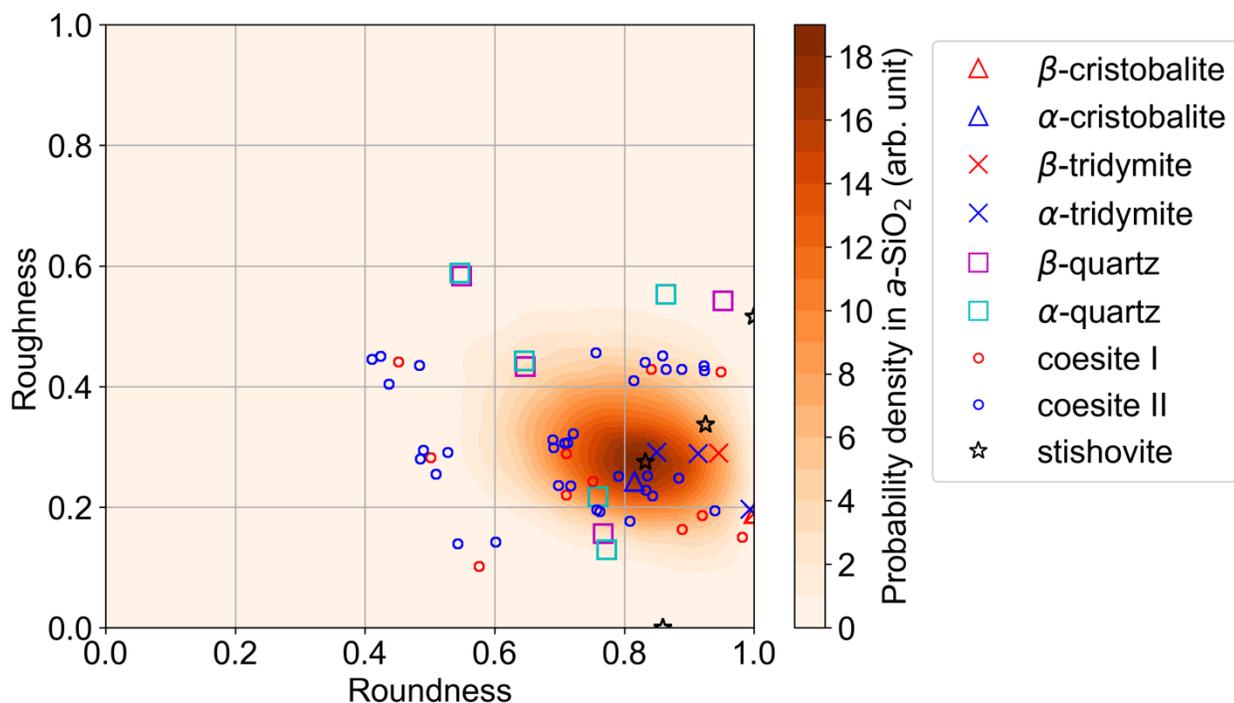

**Fig. 6. Probability distribution of ring characteristic indicators (roundness and roughness) of amorphous and crystalline SiO₂.** The probability density of *a*-SiO$_2$ was computed by a kernel density estimation with Gaussian kernel. The band width of the kernel was determined using the Scott's rule[42] ($=n^{-1/(d+4)}$), where *n* is the number of rings, and *d* is dimension. In the data, *n*=45,423 and *d*=2.

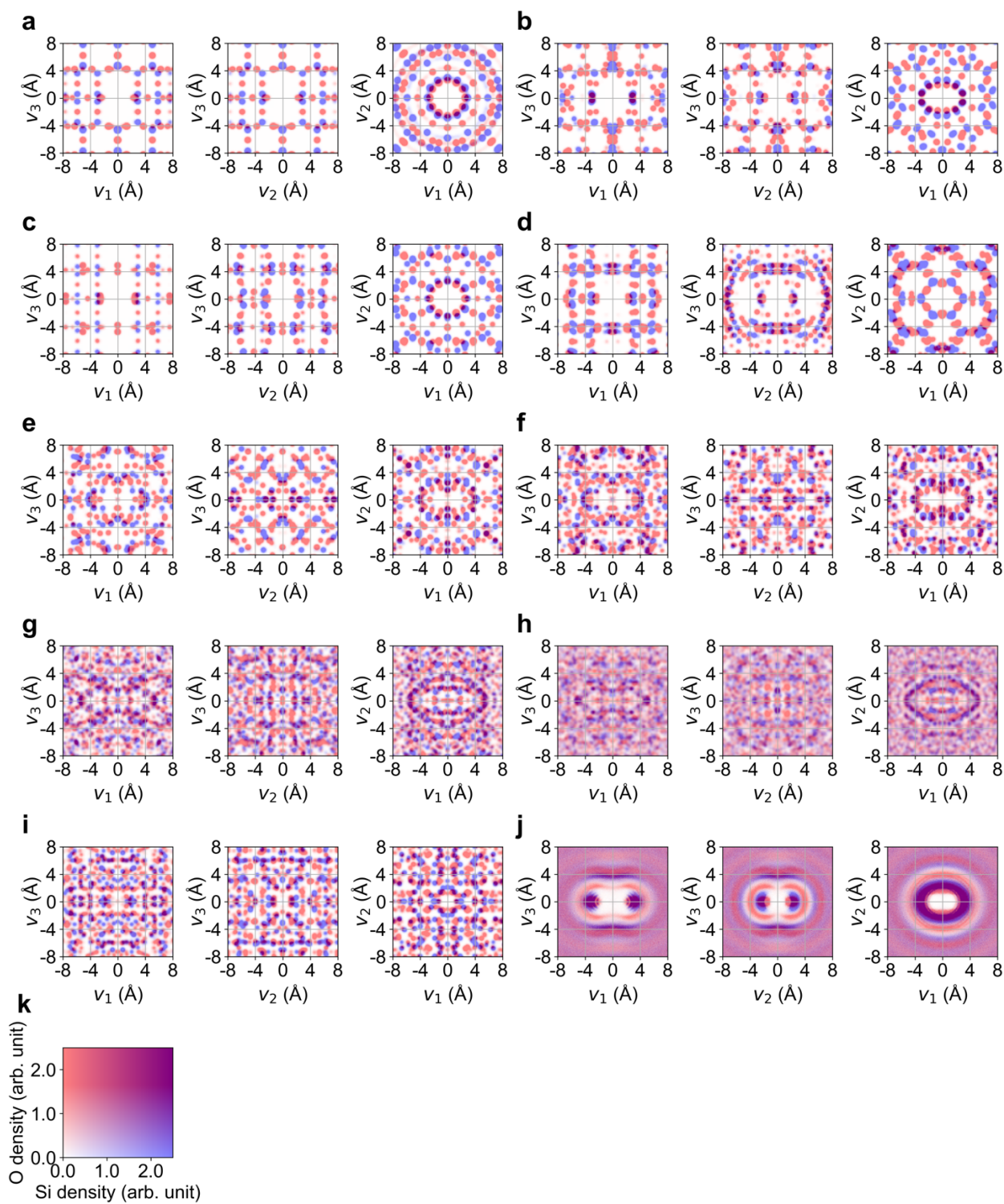

**Fig. 7. Cross-sectional mappings of spatial correlation functions from $v_1$, $v_2$, and $v_3$-axis directions for crystalline and amorphous SiO$_2$.**

**a**–**f** Cross-sectional mappings of the spatial correlation functions of β-cristobalite (**a**), α-cristobalite (**b**), β-tridymite (**c**), α-tridymite (**d**), β-quartz (**e**), α-quartz (**f**), coesite I (**g**), coesite II (**h**), stishovite (**i**), and $a$-SiO$_2$ (MD-RMC) (**j**). **k** Color indicator: a region with blue/red color indicates a large density of Si/O atoms. The cross-sectional thickness $t$ was set to 2 Å for all models.

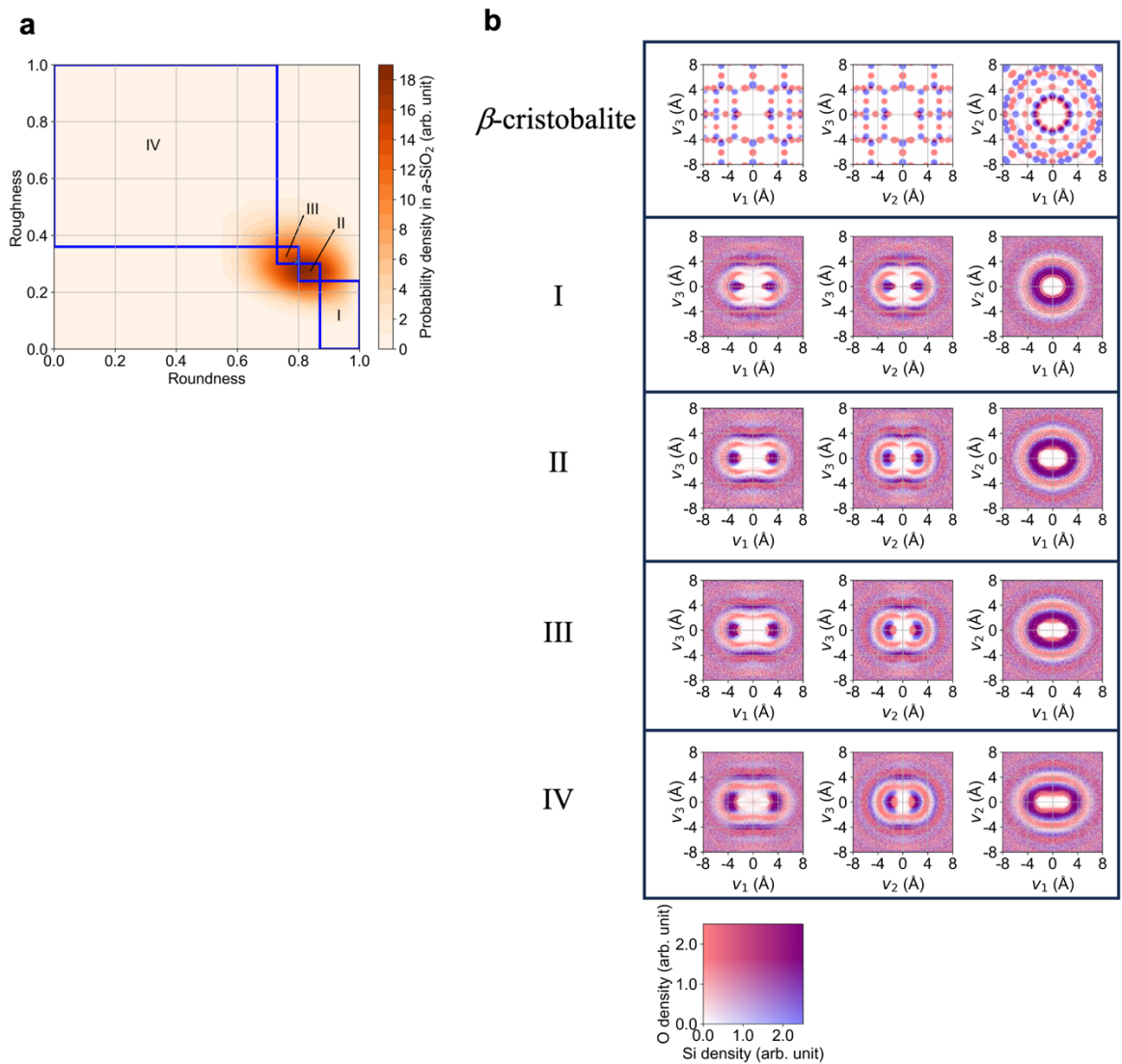

**Fig. 8. Spatial correlations cross-sectional mappings of Si/O atoms using specific shapes of rings in *a*-SiO$_2$. a** Regions of roundness and roughness for computing spatial correlation functions around specific rings. The probability distribution of ring characteristic indicators of amorphous SiO$_2$ is the same as that in Fig. 6. The regions are determined by the quantiles of roundness $r_c = 0.73, 0.80, 0.87$, and those of roughness $r_t = 0.24, 0.30, 0.36$. For example, the range of region I is in $0.87 < r_c \leq 1$ and $0 \leq r_t \leq 0.24$. Rings in region I, whose roundness is larger, and roughness is smaller than those in other regions, are more symmetric than those in other three regions. On the contrary, rings in region IV are less symmetric than others. Region II includes rings of major shapes because the region includes the mode, i.e., the point with the largest probability density. **b** The correlation functions computed using only rings whose shapes are in a specific region (I–IV), together with that of $\beta$-cristobalite, which is same as that in **Fig. 7a**. The color indicator at the bottom shows values of spatial correlations of Si/O atom. The direct comparisons of the function of Si/O atom in each region with that of $\beta$-cristobalite are shown in **Supplementary Figure 13**.

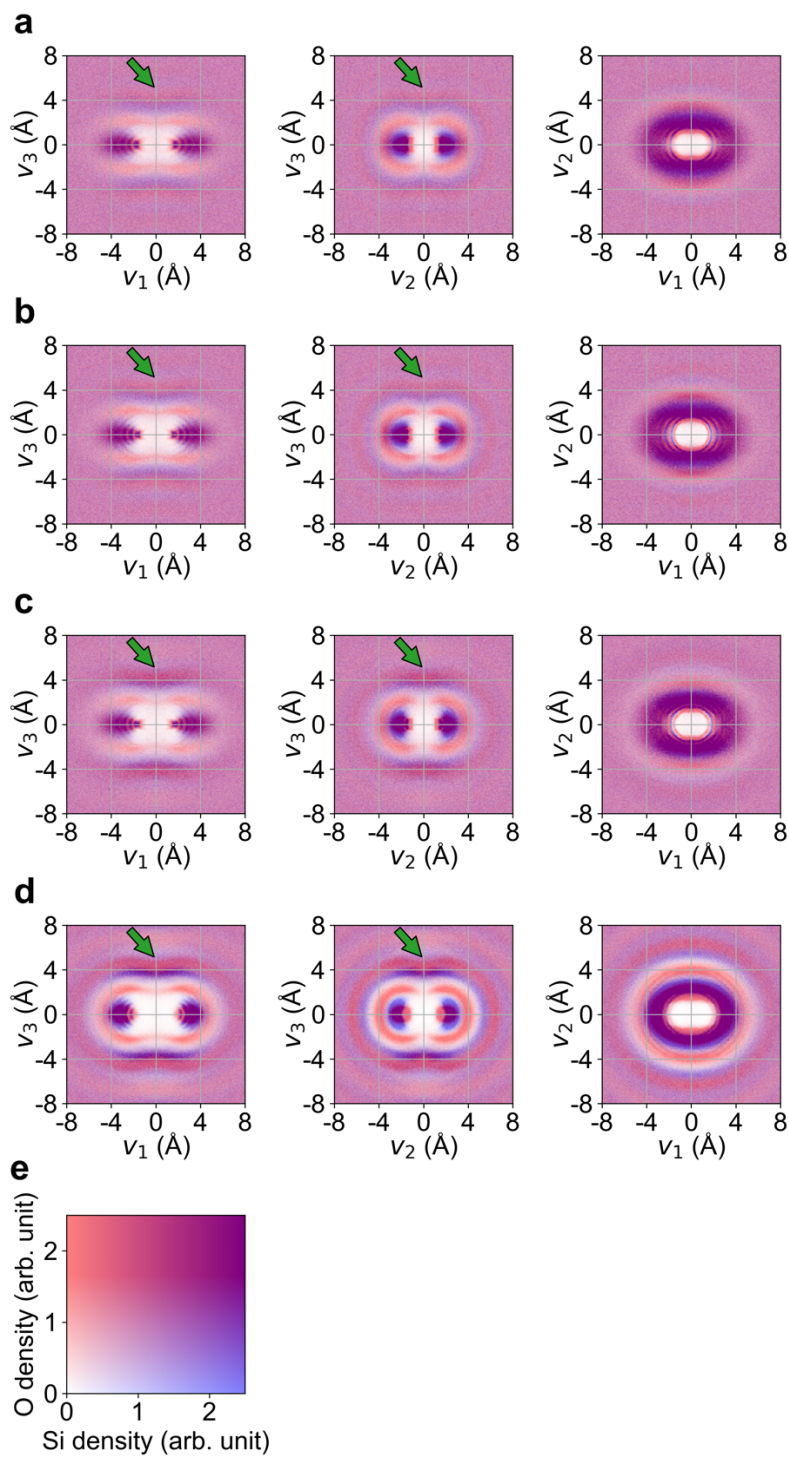

**Fig. 9. Cross-sectional mappings of spatial correlations of *a*-SiO$_2$ models.**

**a**–**d** Cross-sectional mappings of spatial correlations of Rand-Coord (**a**), Rand-Tetra (**b**), Rand-RMC (**c**), and MD-RMC (**d**) models. **e** Color indicator: a region with blue/red color indicates a large density of Si/O atoms. The cross-sectional thickness $t$ is 2 Å. Green arrows indicate parallel planes above and below the rings along the $v_1$ and $v_2$ axes.

# Supplementary Information for

## Ring-originated anisotropy of local structural ordering in amorphous and crystalline silicon dioxide

Motoki Shiga[1,2,3,*,†], Akihiko Hirata[4,5,6], Yohei Onodera[6,7], Hirokazu Masai[8]

*Corresponding author
†motoki.shiga.b4@tohoku.ac.jp

[1] Unprecedented-scale Data Analytics Center, Tohoku University, 468-1 Aoba, Aramaki-Aza, Aoba-ku, Sendai 980-8578, Japan

[2] Graduate School of Information Science, Tohoku University, 6-3-09 Aoba, Aramaki-aza Aoba-ku, Sendai, 980-8579, Japan

[3] RIKEN Center for Advanced Intelligence Project, 1-4-1 Nihonbashi, Chuo-ku, Tokyo 103-0027, Japan

[4] Department of Materials Science, Waseda University, 3-4-1 Ohkubo, Shinjuku, Tokyo 169-8555, Japan.

[5] Kagami Memorial Research Institute for Materials Science and Technology, Waseda University, 2-8-26 Nishiwaseda, Shinjuku, Tokyo 169-0051, Japan

[6] Center for Basic Research on Materials, National Institute for Materials Science, 1-2-1, Sengen, Tsukuba, Ibaraki 305-0047, Japan

[7] Institute for Integrated Radiation and Nuclear Science, Kyoto University, 2-1010 Asashiro-nishi, Kumatori-cho, Sennan-gun, Osaka 590-0494, Japan

[8] National Institute of Advanced Industrial Science and Technology, 1-8-31 Midorigaoka, Ikeda, Osaka 563-8577, Japan

**Supplementary Discussion**

**Structure statistics of crystalline and amorphous model**

Densities and data IDs of the crystalline $SiO_2$ and amorphous models are summarized in **Supplementary Tables 1** and **4**. In all crystalline models, the coordination number of all Si atoms is 4, whereas that of all O atoms is 2. In the amorphous MD-RMC model, the coordination number of 99.8% of Si atoms is 4, whereas that of 99.9% of O atoms is 2. **Supplementary Fig. 1** shows the bond angle distributions of (a) Si–O–Si, (b) O–Si–O, and (c) Si–Si–Si. These figures suggest that the angles of the O–Si–O bonds in all crystalline and amorphous models are concentrated at approximately $109.47° \left(= -\frac{1}{3}\right)$, indicating that the $SiO_4$ tetrahedra in all models are nearly equal to those of a regular tetrahedron. The Si–O–Si bond angles in the low-density structures, such as $\beta$-cristobalite, $\beta$-tridymite, and $\alpha$-tridymite, include a 180° angle, which is linear. Meanwhile, $a$-$SiO_2$, in which the mass density is close to that of these three crystals, also has many such bonds, as indicated by the left-tailed distribution. This suggests that $a$-$SiO_2$ has a structure similar to those of $\beta$-cristobalite, $\beta$-tridymite, and $\alpha$-tridymite. The distributions of Si–O–Si shift to smaller values with increasing mass densities in the crystalline models. Note that coesite I and II, which have the highest densities among crystalline $SiO_2$ materials except for stishovite, have Si–O–Si bonds with an angle of 180°, whereas $\alpha$-cristobalite, $\beta$-quartz, and $\alpha$-quartz do not contain such a bond. In contrast, the variance in the angles of the Si–Si–Si bond increases with the mass densities for crystalline models. This is inversely proportional to the structural order because it inversely reflects the symmetry of the $SiSi_4$ tetrahedra.

The X-ray and neutron total structure factors $S(q)$ of $a$-SiO$_2$ are shown in **Supplementary Figures 2** and **3** together with the results of the MD–RMC model. The MD–RMC model agree well with the experimental data. The coordination numbers around Si and O atoms, of over 99% in the first coordination distance of the generated MD-RMC model, are four and two, respectively. It is confirmed that the MD-RMC model forms a network structure, sharing O atoms at the corners of the SiO$_4$ tetrahedra. The bond angle distributions of the crystalline and amorphous models of MD-RMC are shown in the bottom panels of **Supplementary Figure 1**. The distribution of O–Si–O for $a$-SiO$_2$ (MD-RMC) has a peak around 109.47° $\left(=-\frac{1}{3}\right)$, indicating that the SiO$_4$ tetrahedra are nearly regular. Because the distribution of Si–Si–Si has a peak at approximately 109.47°, the SiSi$_4$ tetrahedra are also close to being regular.

**Structural statistics of the MD model and MD-RMC model**

Because it is hard to construct a reasonable model with only RMC simulations or only MD simulations, RMC was implemented with the constraint of the coordination numbers, bond angles of O–Si–O, and the partial pair-distribution functions within the first coordination shell generated by the MD simulation. As shown in **Supplementary Figures 15** and **16**, coordination numbers and bond angles generated by Rand-RMC, which does not use a MD simulation, are more disordered than those by the MD-RMC model. Meanwhile, only MD simulations make it difficult to fully reproduce the experimental $S(q)$. Therefore, our procedure, wherein MD simulation is followed by RMC, would be one of the best methods for generating a reasonable model. Here, we discuss the differences between the MD-RMC model and MD model.

**Supplementary Figures 2** and **3** show a comparison among the $S(q)$ functions of an experimental measurement, the structure model generated by MD-RMC and that by only MD simulation (MD model), via X-ray diffraction and neutron diffraction. This figure shows that the FSDP by the MD model is considerably sharper than that in the experimental observation. On the contrary, the $S(q)$ of the MD-RMC model is consistent with the experimental one, which is significantly better than that of the MD model over observed $q$ ranges. To clarify these differences, the difference $S(q)$ data is also shown by dashed lines in **Supplementary Figures 2 and 3**. It shows that not only in the FDSP region, but also in the smaller $q$ region below FSDP, the difference by RMC is substantially smaller than that by MD. **Supplementary Figures 4** and **5** show the distributions of coordination numbers and bond angles of the MD model and MD-RMC model. As shown in these figures, structures of $SiO_4$ tetrahedra generated by the MD simulation are retained, whereas the distribution of Si–O–Si and Si–Si–Si angles in **Supplementary Figures 5a** and **5b**, which shows the measurement of inter-tetrahedral structures, are changed to fit the experimental structure factor. Then, RMC refines the data to match the experimental $S(q)$ while maintaining the basic structure, i.e., with structural constraints. As mentioned above, the RMC refinement did not violate structures in the first coordinates generated by the MD simulation.

**Structure statistics of amorphous models with different structural restrictions**

**Supplementary Figure 14** shows structure factors by X-ray diffraction, (**Supplementary Figure 14a**) neutron diffraction (**Supplementary Figure 14b**), and partial structure factors (**Supplementary Figures 14c–e**) of four amorphous models: Rand-Coord, Rand-Tetra, Rand-RMC, and MD-RMC. **Supplementary Figures 14a–b** demonstrate that Rand-RMC and MD-RMC exhibit FSDP at around $q = 1.5\ \text{Å}^{-1}$, which is not the case for Rand-Coord and Rand-Tetra. In **Supplementary Figures 14c–e**, partial structure factors of only MD-RMD model have sharp principal peaks (PPs) at around $q = 2.8\ \text{Å}^{-1}$.

In addition, **Supplementary Figure 15** shows coordination number distributions of the four models. In random initialization models (i.e., Rand-Coord, Rand-Tetra, and Rand-RMC), the coordination numbers of only approximately 95% of Si atoms are four, while those of approximately 97% of O atoms are two. These percentages are smaller than those in the MD-RMC model. Furthermore, **Supplementary Figures 16** and **17** show bond angle distributions and ring characterizations, respectively. These figures show that the random initialization models are more broadly distributed than those in the MD-RMC model. In particular, the distributions of the ring sizes of the random initialization models are considerably different from those of the MD-RMC model. Overall, these structural statistics demonstrate that among all the models, the MD-RMC model is the most structurally ordered.

**Supplementary Table 1. Atomic density of crystalline SiO$_2$.** AMCSD: American Mineralogist Crystal Structure Database, COD: Crystallography Open Database, ICSD: Inorganic Crystal Structure Database, and SM: Springer Materials.

| Name | Mass density (g cm$^{-3}$) | Number density (#atoms Å$^{-3}$) | Database and ID |
| --- | --- | --- | --- |
| $\beta$-cristobalite | 2.21 | 0.0665 | AMCSD #0017665 |
| $\alpha$-cristobalite | 2.33 | 0.0701 | COD #9009686 |
| $\beta$-tridymite | 2.22 | 0.0666 | COD #5910147 |
| $\alpha$-tridymite | 2.21 | 0.0665 | ICSD #40895 |
| $\beta$-quartz | 2.54 | 0.0763 | SM #sd_1144624 |
| $\alpha$-quartz | 2.65 | 0.0796 | COD #9005017 |
| coesite I | 2.92 | 0.0878 | COD #9000803 |
| coesite II | 3.68 | 0.1108 | SM #sd_1722202 |
| stishovite | 4.28 | 0.1287 | SM #sd_1938143 |

**Supplementary Table 2. Symmetry information of crystalline SiO$_2$.** Point group is described by Hermann–Mauguin notation.

| Name | Space group | Point group | Crystal system |
| --- | --- | --- | --- |
| $\beta$-cristobalite | F d $\bar{3}$ m | m $\bar{3}$ m | Cubic |
| $\alpha$-cristobalite | P 4$_1$ 2$_1$ 2 | 4 2 2 | Tetragonal |
| $\beta$-tridymite | P 6$_3$/m m c | 6/m m m | Hexagonal |
| $\alpha$-tridymite | C 2 2 2$_1$ | 2 2 2 | Orthorhombic |
| $\beta$-quartz | P 6$_2$ 2 2 | 6 2 2 | Hexagonal |
| $\alpha$-quartz | P 3$_2$ 2 1 | 3 2 | Trigonal |
| coesite I | C 2/c | 2/m | Monoclinic |
| coesite II | P 2$_1$/c | 2/m | Monoclinic |
| stishovite | P 4$_2$/m n m | 4/m m m | Tetragonal |

**Supplementary Table 3. Bravais lattices of crystalline SiO$_2$.**

| Name | length ($a$, $b$, $c$) (Å) | angle ($\alpha$, $\beta$, $\gamma$) (deg.) | Cell volume (Å$^3$) |
|---|---|---|---|
| $\beta$-cristobalite | 7.120, 7.120, 7.120 | 90.0, 90.0, 90.0 | 360.944 |
| $\alpha$-cristobalite | 4.971, 4.971, 6.928 | 90.0, 90.0, 90.0 | 171.185 |
| $\beta$-tridymite | 5.030, 5.030, 8.220 | 90.0, 90.0, 120.0 | 180.110 |
| $\alpha$-tridymite | 8.756, 5.024, 8.213 | 90.0, 90.0, 90.0 | 361.291 |
| $\beta$-quartz | 4.996, 4.996, 5.457 | 90.0, 90.0, 120.0 | 117.967 |
| $\alpha$-quartz | 4.914, 4.914, 5.405 | 90.0, 90.0, 120.0 | 113.011 |
| coesite I | 7.136, 12.369, 7.174 | 90.0, 120.34, 90.0 | 546.439 |
| coesite II | 6.545, 23.039, 6.503 | 90.0, 117.87, 90.0 | 866.746 |
| stishovite | 4.181, 4.181, 2.666 | 90.0, 90.0, 90.0 | 46.612 |

**Supplementary Table 4. Statistics on the atoms in $a$-SiO$_2$ models (MD-RMC, Rand-Coord, Rand-Tetra, and Rand-RMC).**

| Name | Mass density (g cm$^{-3}$) | Number density (#atoms Å$^{-3}$) | Length of simulation box (Å) | Number of Si atoms | Number of O atoms |
|---|---|---|---|---|---|
| $a$-SiO$_2$ | 2.20 | 0.0662 | 100.0 | 22,052 | 44,104 |

**Supplementary Table 5. Shape characters of isomorphic rings in crystalline $SiO_2$.** IDs are identical to those indicated in Fig. 4. Point group is described by Hermann–Mauguin notation.

| Name | ID | Ring size | Roundness | Roughness | Point group |
|---|---|---|---|---|---|
| β-cristobalite | A6-1 | 6 | 1.00 | 0.19 | $\bar{3}$ m |
| α-cristobalite | B6-1 | 6 | 0.82 | 0.24 | 1 |
| β-tridymite | C6-1 | 6 | 1.00 | 0.19 | $\bar{3}$ m |
| | C6-2 | 6 | 0.95 | 0.29 | m m 2 |
| α-tridymite | D6-1 | 6 | 0.99 | 0.20 | 2 |
| | D6-2 | 6 | 0.91 | 0.29 | 1 |
| | D6-3 | 6 | 0.85 | 0.29 | 2 |
| β-quartz | E6-1 | 6 | 0.95 | 0.54 | 2 2 2 |
| | E8-1 | 8 | 0.77 | 0.16 | 2 |
| | E8-2 | 8 | 0.65 | 0.43 | 2 |
| | E8-3 | 8 | 0.55 | 0.58 | 2 2 2 |
| α-quartz | F6-1 | 6 | 0.86 | 0.55 | 2 |
| | F8-1 | 8 | 0.77 | 0.13 | 2 |
| | F8-2 | 8 | 0.76 | 0.22 | 2 |
| | F8-3 | 8 | 0.65 | 0.44 | 1 |
| | F8-4 | 8 | 0.55 | 0.59 | 2 |
| coesite I | G4-1 | 4 | 0.92 | 0.19 | 2 |
| | G4-2 | 4 | 0.89 | 0.16 | $\bar{1}$ |
| | G6-1 | 6 | 0.75 | 0.24 | $\bar{1}$ |
| | G8-1 | 8 | 0.98 | 0.15 | $\bar{1}$ |
| | G9-1 | 9 | 0.50 | 0.28 | 1 |
| | G9-2 | 9 | 0.45 | 0.44 | 1 |
| | G10-1 | 10 | 0.58 | 0.10 | $\bar{1}$ |
| | G11-1 | 11 | 0.95 | 0.42 | 1 |
| | G11-2 | 11 | 0.84 | 0.43 | 1 |
| | G12-1 | 12 | 0.71 | 0.22 | $\bar{1}$ |
| | G12-2 | 12 | 0.71 | 0.29 | 1 |
| coesite II | H4-1 | 4 | 0.84 | 0.22 | 1 |
| | H4-2 | 4 | 0.84 | 0.25 | $\bar{1}$ |
| | H4-3 | 4 | 0.81 | 0.18 | 1 |
| | H4-4 | 4 | 0.76 | 0.19 | $\bar{1}$ |
| | H4-5 | 4 | 0.76 | 0.20 | $\bar{1}$ |
| | H6-1 | 6 | 0.88 | 0.25 | $\bar{1}$ |
| | H6-2 | 6 | 0.72 | 0.32 | $\bar{1}$ |
| | H6-3 | 6 | 0.79 | 0.25 | $\bar{1}$ |
| | H8-1 | 8 | 0.94 | 0.19 | 1 |
| | H8-2 | 8 | 0.83 | 0.23 | 1 |
| | H9-1 | 9 | 0.44 | 0.40 | 1 |
| | H9-2 | 9 | 0.48 | 0.44 | 1 |
| | H9-3 | 9 | 0.49 | 0.28 | 1 |
| | H9-4 | 9 | 0.49 | 0.29 | 1 |

|  | H9-5 | 9 | 0.51 | 0.26 | 1 |
|---|---|---|---|---|---|
|  | H9-6 | 9 | 0.53 | 0.29 | 1 |
|  | H9-7 | 9 | 0.41 | 0.45 | 1 |
|  | H9-8 | 9 | 0.42 | 0.45 | 1 |
|  | H10-1 | 10 | 0.54 | 0.14 | 1 |
|  | H10-2 | 10 | 0.60 | 0.14 | 1 |
|  | H11-1 | 11 | 0.92 | 0.43 | 1 |
|  | H11-2 | 11 | 0.92 | 0.43 | 1 |
|  | H11-3 | 11 | 0.89 | 0.43 | 1 |
|  | H11-4 | 11 | 0.86 | 0.43 | 1 |
|  | H11-5 | 11 | 0.86 | 0.45 | 1 |
|  | H11-6 | 11 | 0.83 | 0.44 | 1 |
|  | H11-7 | 11 | 0.81 | 0.41 | 1 |
|  | H11-8 | 11 | 0.76 | 0.46 | 1 |
|  | H12-1 | 12 | 0.69 | 0.30 | 1 |
|  | H12-2 | 12 | 0.70 | 0.24 | 1 |
|  | H12-3 | 12 | 0.71 | 0.31 | 1 |
|  | H12-4 | 12 | 0.71 | 0.31 | 1 |
|  | H12-5 | 12 | 0.72 | 0.24 | 1 |
|  | H12-6 | 12 | 0.69 | 0.31 | 1 |
| stishovite | I2-1 | 2 | 0.86 | 0.00 | m m m |
|  | I3-1 | 3 | 0.93 | 0.34 | m |
|  | I4-1 | 4 | 1.00 | 0.52 | $\bar{4}$ |
|  | I4-2 | 4 | 0.83 | 0.28 | $\bar{1}$ |

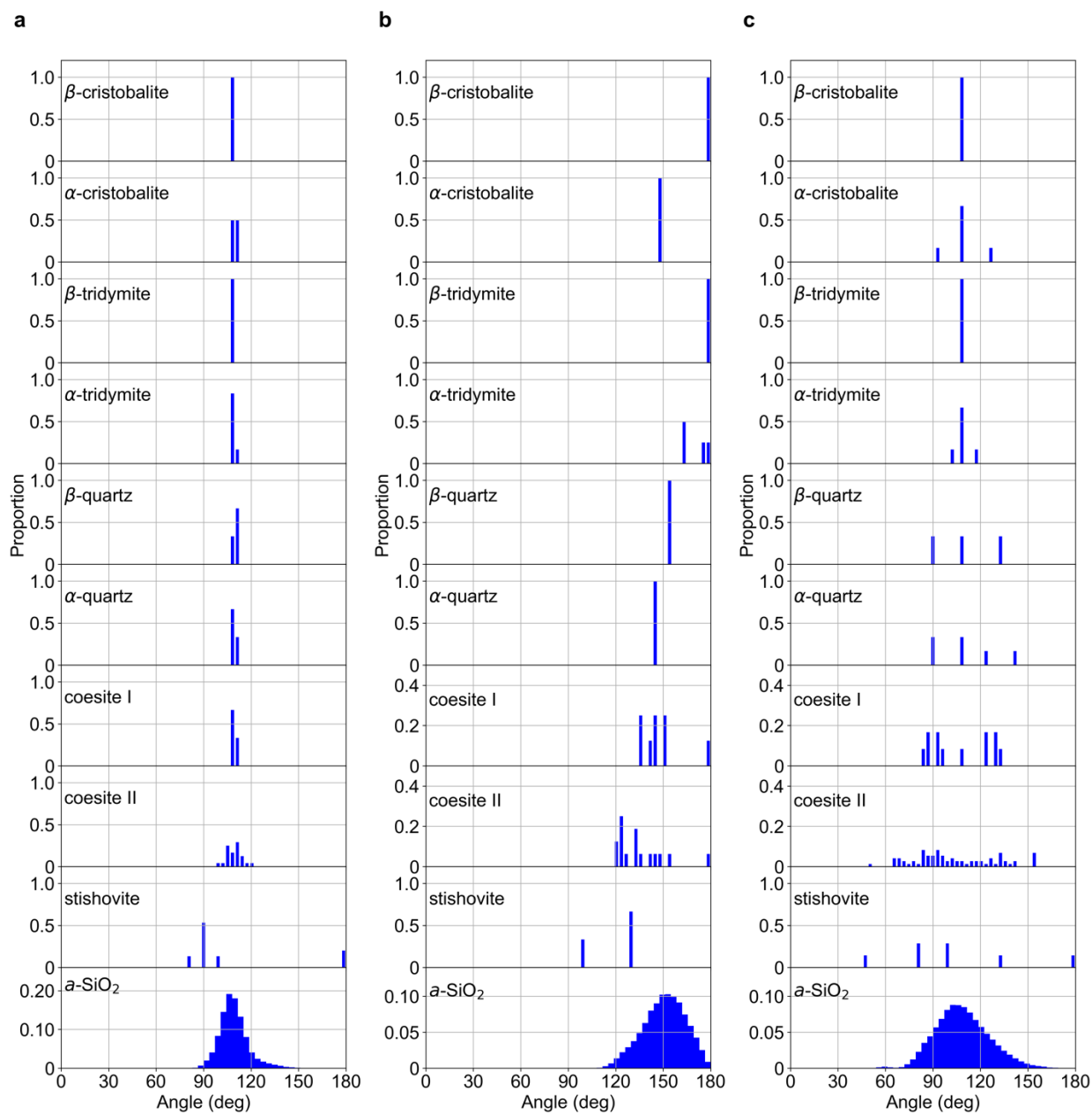

**Supplementary Figure 1. Triplet correlations evaluated by bond angle distribution in crystalline and amorphous SiO₂.** Bond angle distribution of O–Si–O (**a**), Si–O–Si (**b**), and Si–Si–Si (**c**).

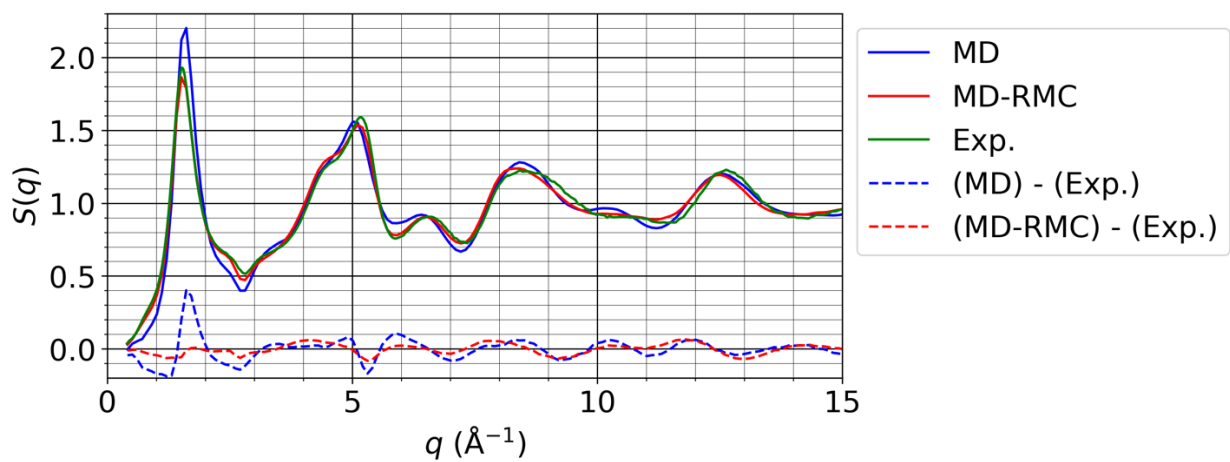

**Supplementary Figure 2. Total structure factors $S(q)$ obtained via X-ray diffraction for $a$-SiO$_2$ models (MD and MD-RMC) and the experimental data.** The difference between the X-ray diffraction data computed from a structural model and that by the experiment is shown by a dashed line.

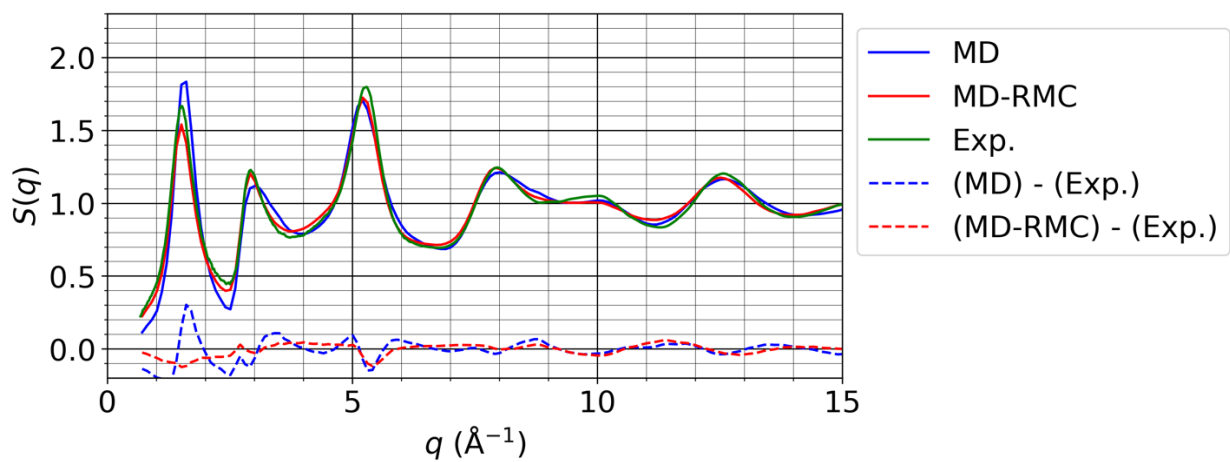

**Supplementary Figure 3. Total structure factors $S(q)$ obtained via Neutron diffraction data for $a$-SiO$_2$ models (MD and MD-RMC) and the experimental data.** The difference between the neutron diffraction data computed from a structural model and that by the experiment is shown by a dashed line.

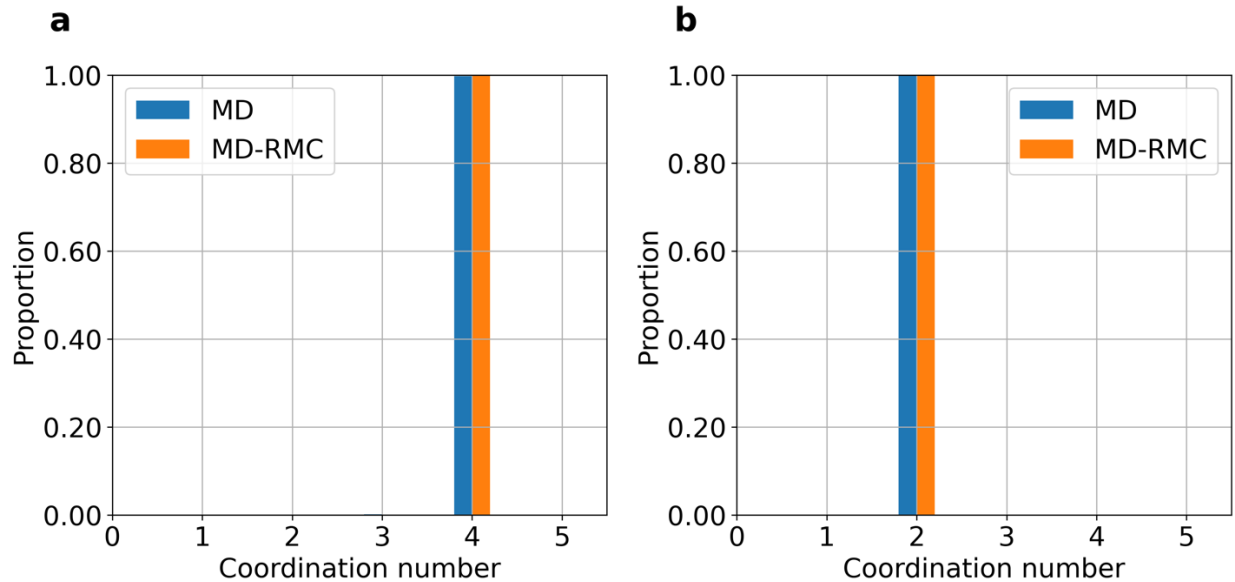

**Supplementary Figure 4. Distribution of coordination numbers in *a*-SiO$_2$ models: MD and MD-RMC.** Coordination numbers of Si atom (**a**) and O atom (**b**).

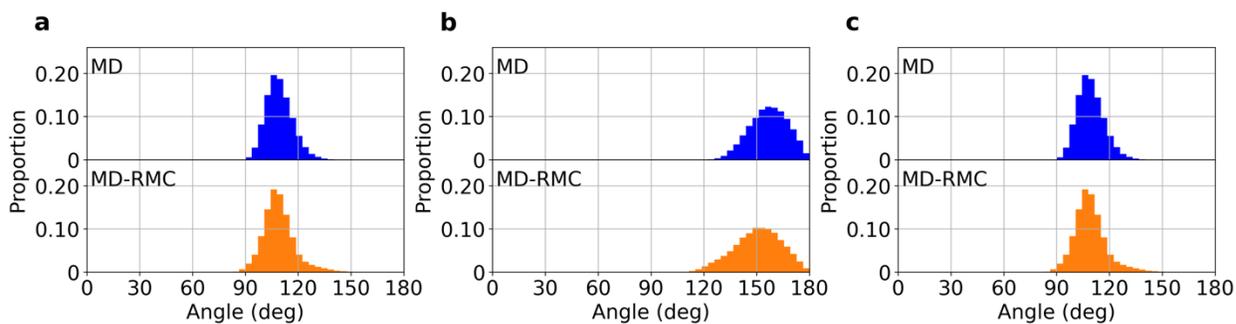

**Supplementary Figure 5. Triplet correlations evaluated by bond angle distributions in *a*-SiO$_2$ models: MD and MD-RMC.** Bond angle distribution of O–Si–O (**a**), Si–O–Si (**b**), and Si–Si–Si (**c**).

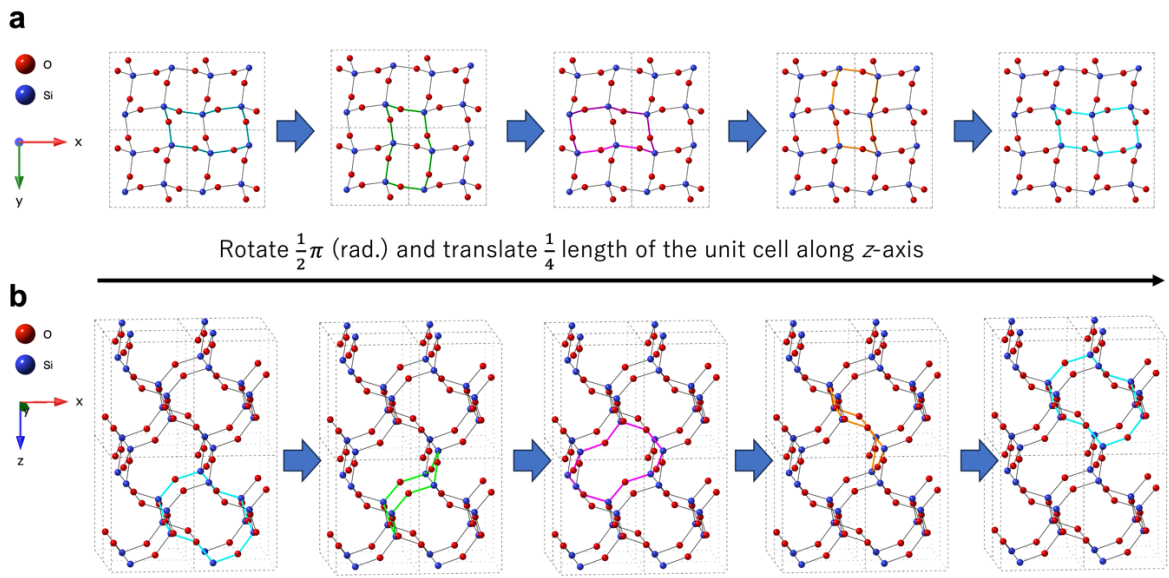

**Supplementary Figure 6. Symmetry in α-cristobalite. a** The view from the direction [001] of α-cristobalite. **b** The side view of that in (**a**). The operation of the screw symmetry along [001] is the combination of a rotation of $1/2\,\pi$ radian and a translation of $1/4$ the lattice length, as shown in these panels. The cyan ring (B6-1) in the left most diagram moves to the green ring in the second diagram by the operation. After four operations, the ring moves to the next cell, which is the same as the parallel translation along the unit cell. In α-cristobalite, the ring shape is not directly linked to the symmetry operation because of the non-symmetric shape of the ring.

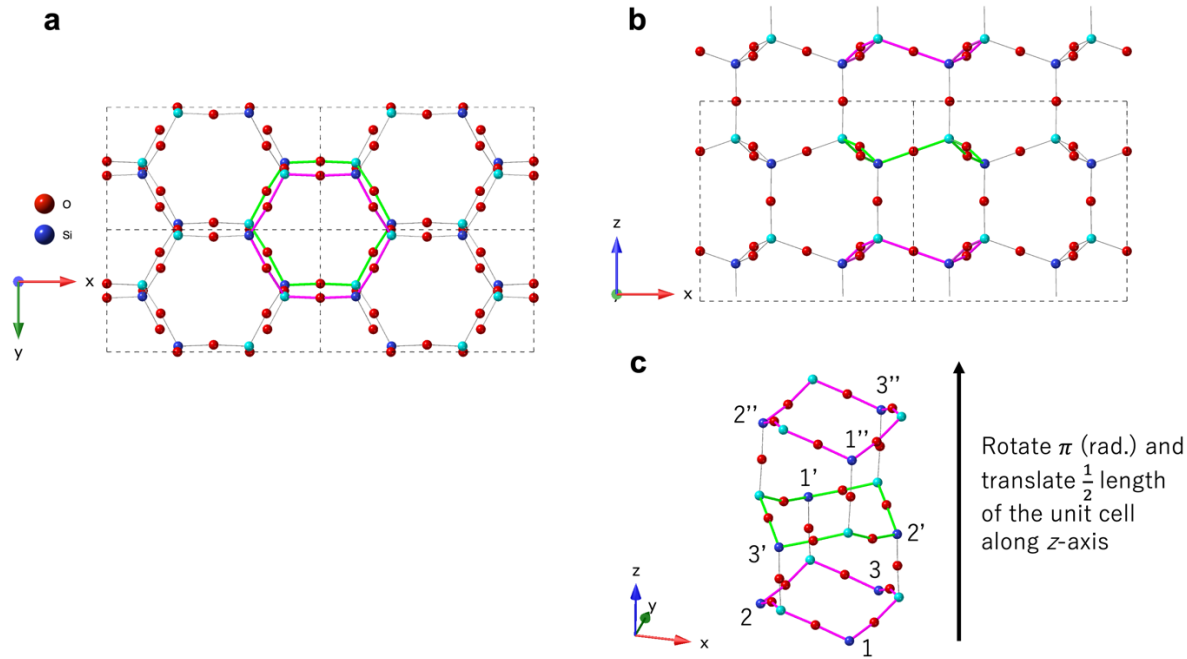

**Supplementary Figure 7. Symmetry in α-tridymite**. **a** The view from the direction [001] of α-tridymite. The magenta ring indicates the ring D6-1. The point group of the ring is 2, whose rotation axis is identical to the second eigenvector of the ring. **b** The view from the direction [010]. Si atoms are colored blue or cyan, which indicate site equivalence whether the $z$ coordinate is lower or higher from the ring plane. **c** The screw symmetry $2_1$ along the direction [001]. Numbers marked in prime and double primes in the ring D6-1 indicate trajectories of the screw operation. The symmetry is not related to the rotation symmetry of the D6-1 ring because the rotation axis is not identical to the normal vector of the ring.

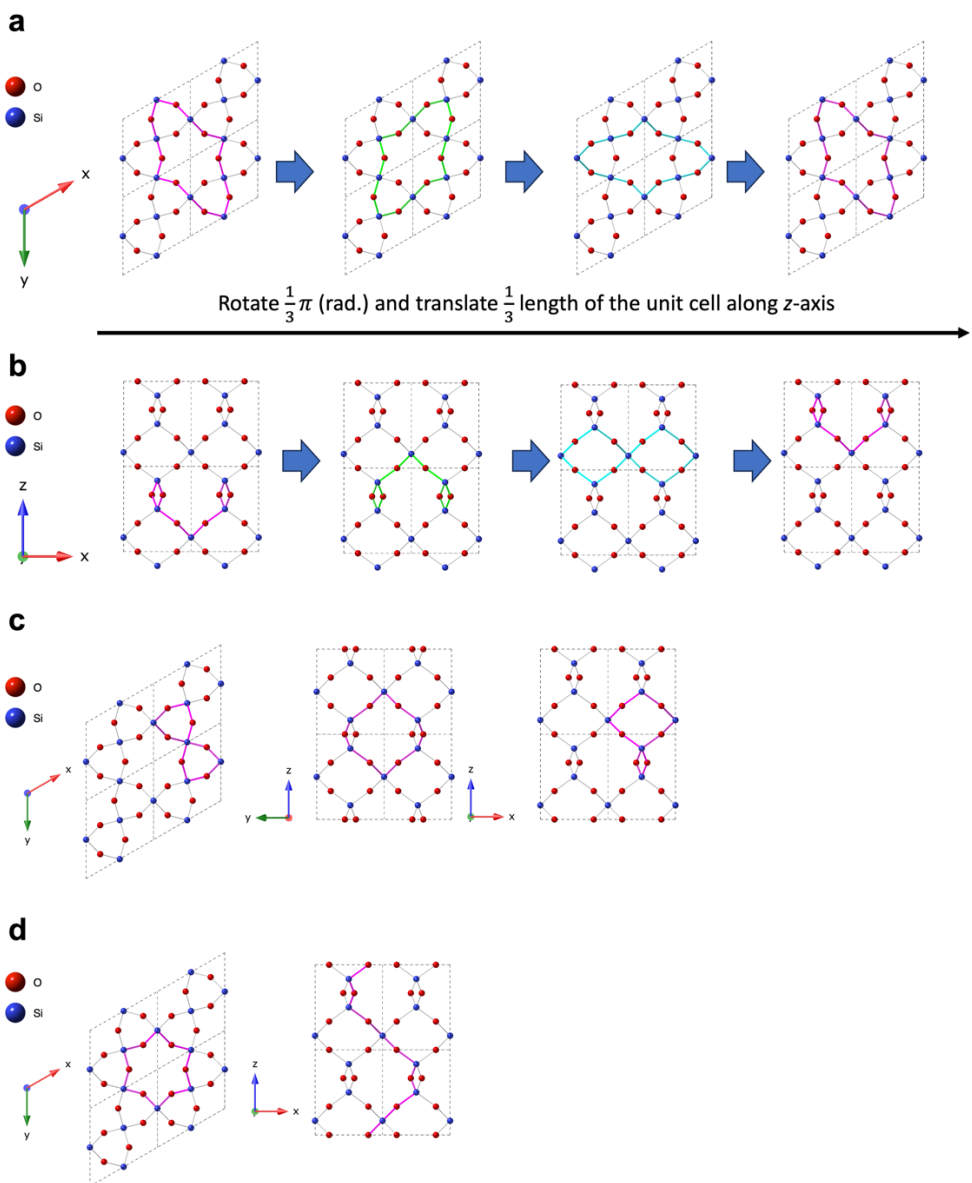

**Supplementary Figure 8. Symmetry with rings in β-quartz. a** The view from the direction [001] of α-quartz. **b** The view from the direction [010]. These panels show the trajectory of the ring E8-3 by the operation of screw symmetry along [001], which is the combination of a rotation of $1/3\pi$ radian and a translation of $1/3$ of the lattice length. The magenta ring in the left most diagram moves to the green ring in the second diagram by the operation. After three operations, the ring moves to the neighbor cell, which is the same as the parallel translation along the unit cell. **c** The views of the six-fold ring (E6-1) from three directions. The ring is not linked to the screw symmetry. **d** The silhouette of the central hole seems to be a ring. However, as shown in the right diagram, the path is not a ring since it is not closed. Therefore, β-quartz does not include any rings linked to the crystal symmetry.

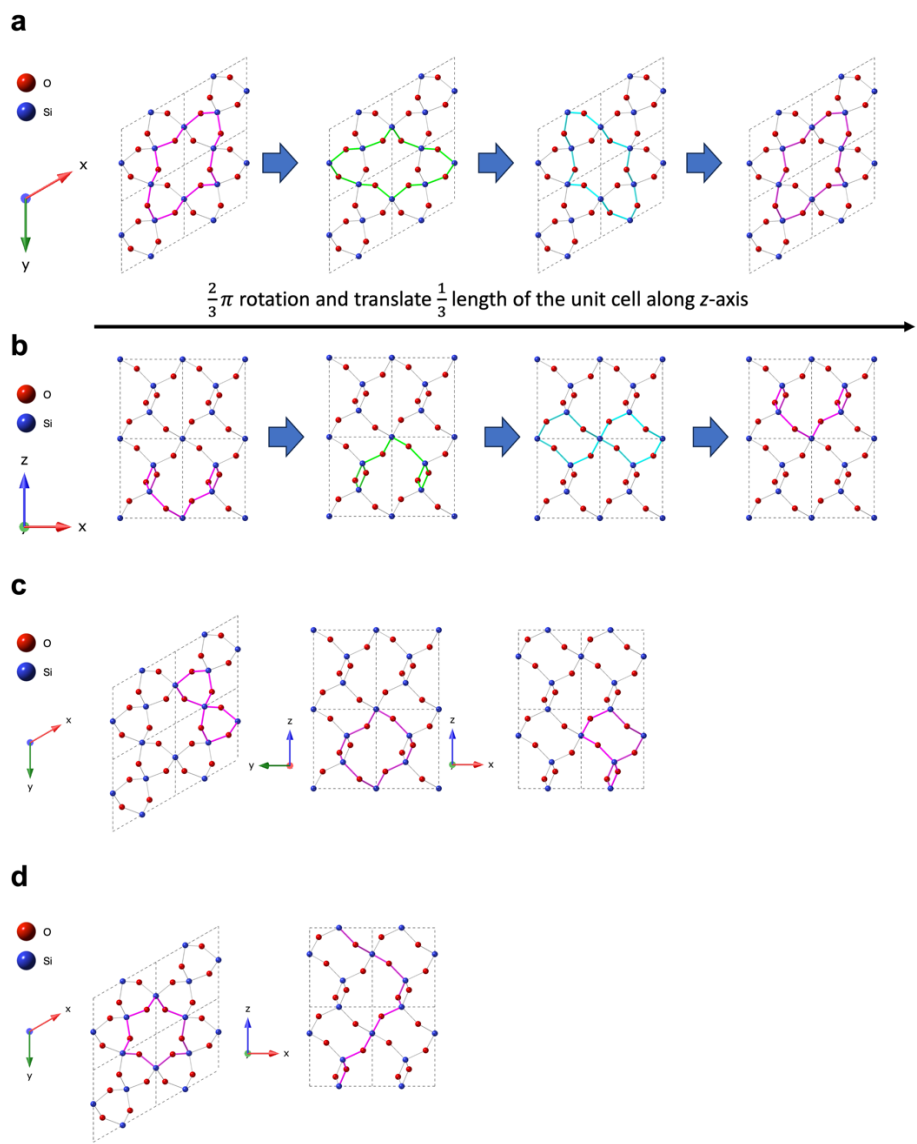

**Supplementary Figure 9. Symmetry with rings in α-quartz**. **a** The view from the direction [001] of β-quartz. **b** The view from the direction [010]. These panels show the trajectory of the ring F8-4 by the operation of screw symmetry along [001], which is the combination of a rotation of $2/3\pi$ radian and a translation of $1/3$ of the lattice length. The magenta ring in the left most diagram moves to the green ring in the second diagram by the operation. After three operations, the ring moves to the neighbor cell, which is same as the parallel translation along the unit cell. **c** The views of the six-fold ring (F6-1) from three directions. The ring is not linked to the screw symmetry. **d** The path related to the silhouette of the central hole, which is less symmetric than that in β-quartz. As shown in the right diagram, the path is not a ring since it not closed. From these observations, α-quartz does not include any rings linked to the crystal symmetry.

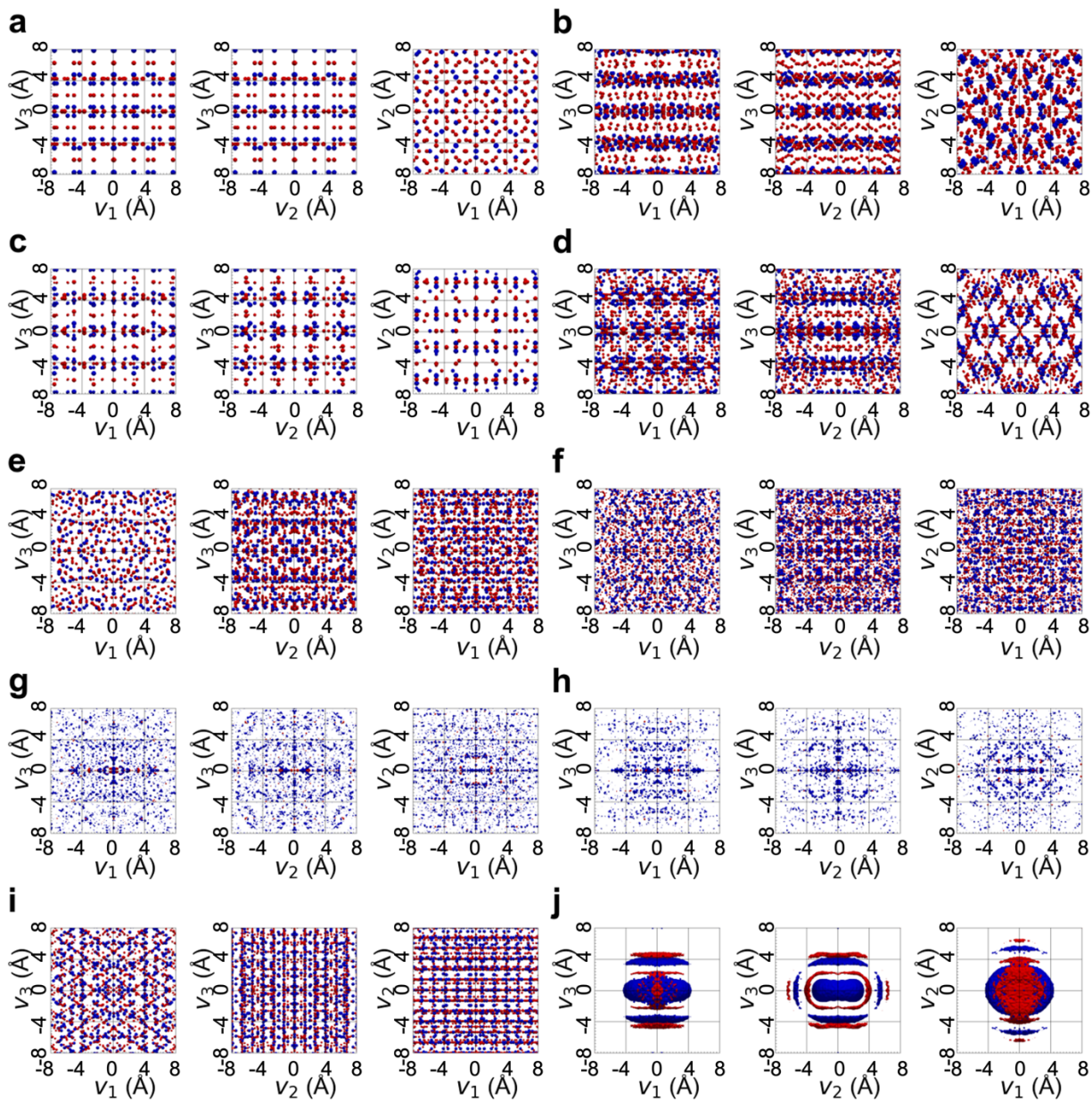

**Supplementary Figure 10. Three-dimensional visualization of surfaces with large correlations along $v_1$-, $v_2$-, and $v_3$-axes for crystalline and amorphous SiO$_2$.** Three-dimensional visualization of spatial correlation functions of $\beta$-cristobalite (**a**), $\alpha$-cristobalite (**b**), $\beta$-tridymite (**c**), $\alpha$-tridymite (**d**), $\beta$-quartz (**e**), $\alpha$-quartz (**f**), coesite I (**g**), coesite II (**h**), stishovite (**i**), and $a$-SiO$_2$ (**j**). In these diagrams, regions whose correlation values are larger than the threshold are visualized in blue for Si atom and red for O atoms. The threshold of correlations for surface visualization for coesite II, other crystals, and amorphous materials are 7, 15, and 1.5, respectively. These threshold values were determined to visualize such regions clearly. Different values were used for amorphous and crystalline materials because of the difference of structural orders.

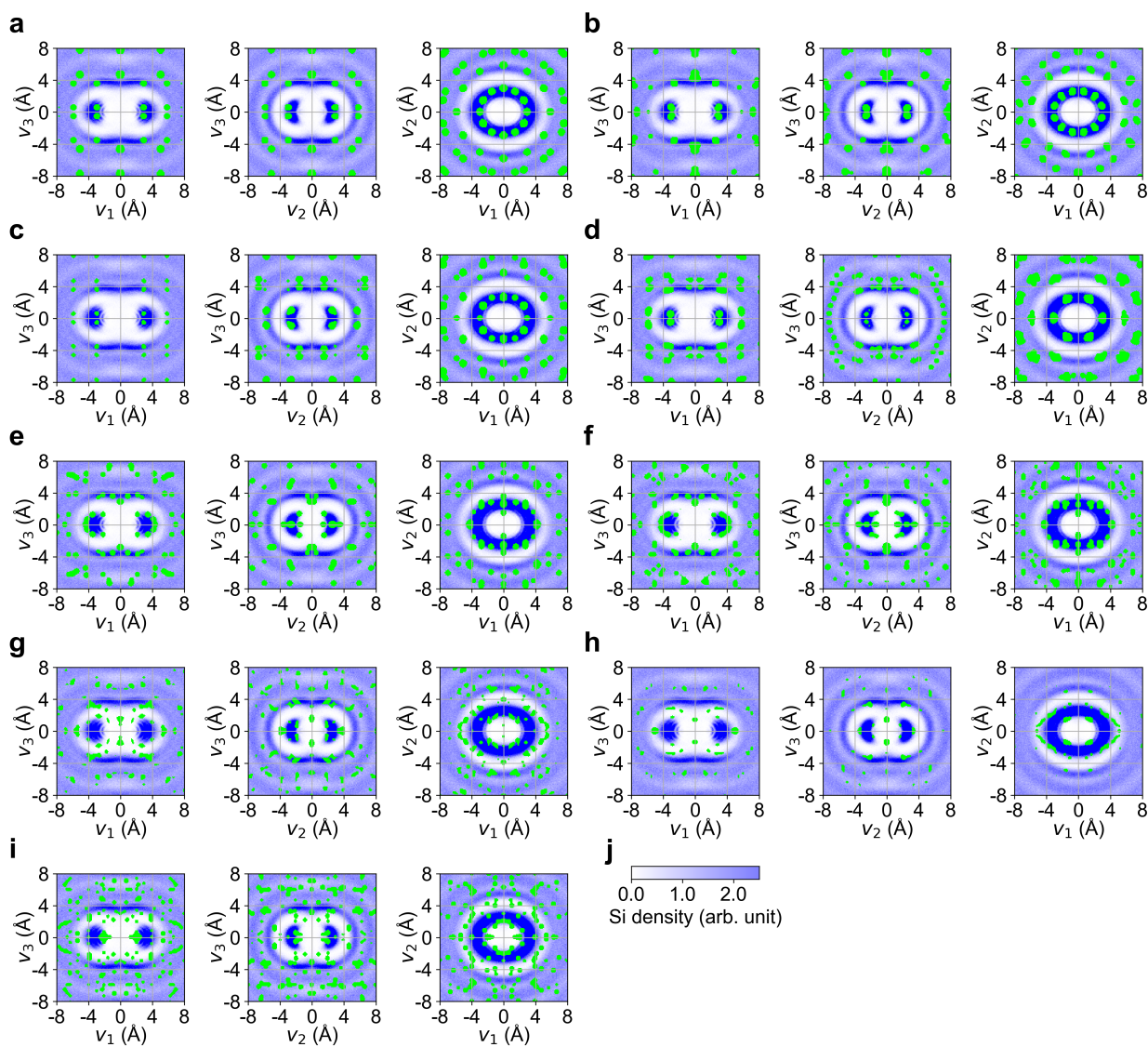

**Supplementary Figure 11. Comparison of the spatial correlations cross-sectional mappings of Si atoms from the $v_1$-, $v_2$-, and $v_3$-axis directions in $a$-SiO$_2$ with those in a crystalline SiO$_2$. a–e** The blue region indicates a spatial correlation of Si atom in $a$-SiO$_2$, whereas the green region indicates spatial positions where the correlation of Si atom is larger than the threshold in a crystalline material: $\beta$-cristobalite (**a**), $\alpha$-cristobalite (**b**), $\beta$-tridymite (**c**), $\alpha$-tridymite (**d**), $\beta$-quartz (**e**), $\alpha$-quartz (**f**), coesite I (**g**), coesite II (**h**), and stishovite (**i**). **j** Color bar indicating the spatial correlation of $a$-SiO$_2$. These diagrams visualize only large correlation regions in functions of crystalline materials, which are colored green, larger than the threshold, which is set to 2.0. The setting of the threshold is not sensitive for visualization because regions with large correlation values and those with small values are clearly separated in the functions of crystalline materials.

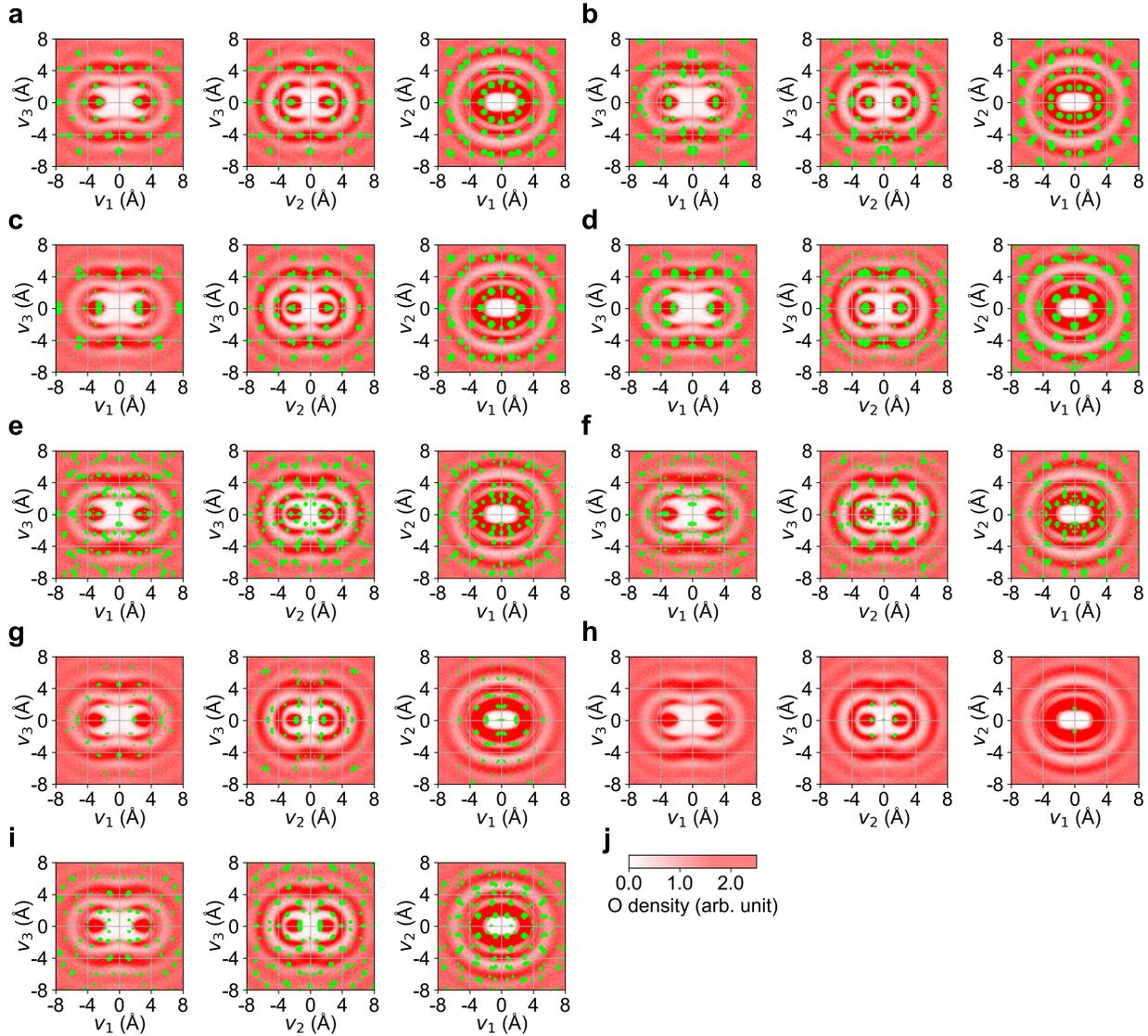

**Supplementary Figure 12. Comparison of the spatial correlations cross-sectional mappings of O atoms from the $v_1$-, $v_2$-, and $v_3$-axis directions in $a$-SiO$_2$ with those in a crystalline SiO$_2$.** **a–e** The red region indicates a spatial correlation of O atom in $a$-SiO$_2$, whereas the green region indicates spatial positions where the correlation of O atom is larger than the threshold in a crystalline material: $\beta$-cristobalite (**a**), $\alpha$-cristobalite (**b**), $\beta$-tridymite (**c**), $\alpha$-tridymite (**d**), $\beta$-quartz (**e**), $\alpha$-quartz (**f**), coesite I (**g**), coesite II (**h**), and stishovite (**i**). **j** Color bar indicating the spatial correlation of $a$-SiO$_2$. These diagrams visualize only large correlation regions in functions of crystalline materials, which are colored green, larger than the threshold, which is set to 2.0. The setting of the threshold is not sensitive for visualization because regions with large correlation values and those with small values are clearly separated in the functions of crystalline materials.

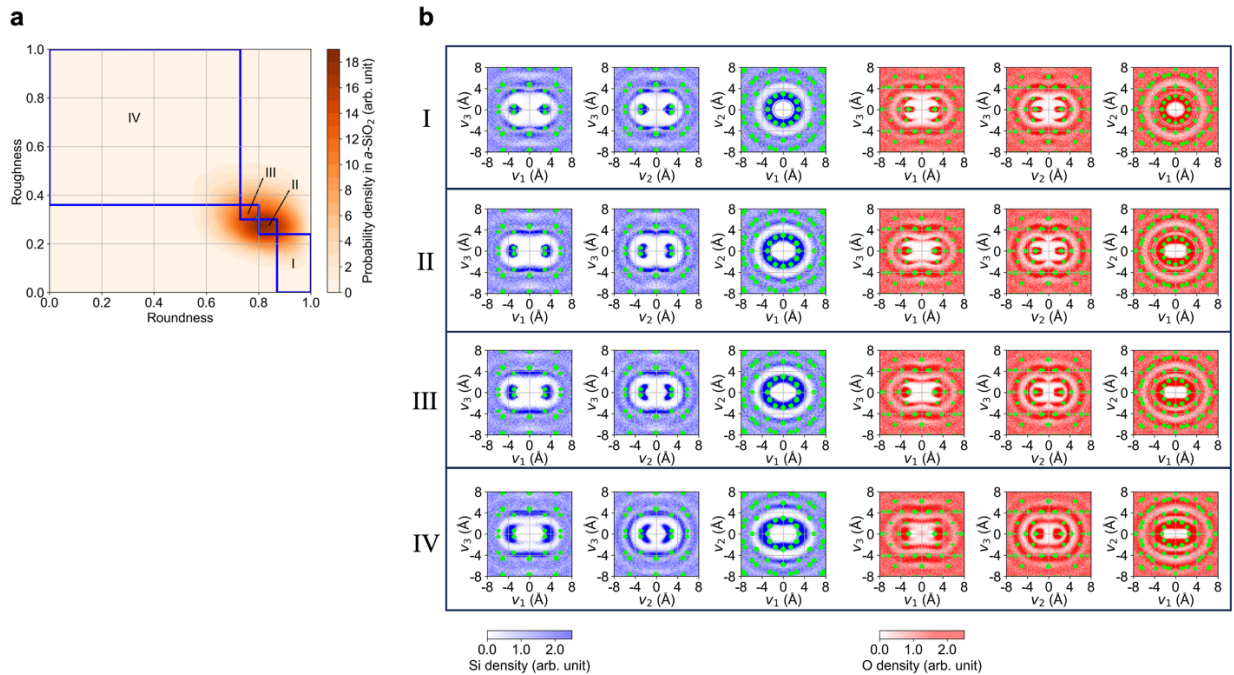

**Supplementary Figure 13. Comparison of spatial correlations cross-sectional mappings of Si/O atoms using specific shapes of rings in *a*-SiO$_2$ with those in *β*-cristobalite. a** Regions of roundness and roughness for computing spatial correlation functions around the specific rings. The probability distribution of ring characteristic indicators of amorphous SiO$_2$, the same as that in Fig. 5. The regions are determined by the quantiles of roundness $r_c = 0.73, 0.80, 0.87$, and those of roughness $r_t = 0.24, 0.30, 0.36$. For example, the range of the region I is $0.87 < r_c \leq 1$ and $0 \leq r_t \leq 0.24$. Rings in region I, whose roundness is larger, and roughness is smaller than those in other regions, are more symmetric than those in other three regions. On the contrary, the rings in region IV are less symmetric than others. Region II includes rings of major shapes because the region includes the mode, i.e., the point with the largest probability density. **b** The correlation functions computed using only rings whose shapes are in a specific region (I–IV). The blue density indicates a spatial correlation of Si atoms in *a*-SiO$_2$ while the red density indicates that of O atoms. The green region indicates the spatial positions in *β*-cristobalite where the correlation of the Si atom is larger than the threshold, which is set to 2.0. Color bars at the bottom indicate the spatial correlation of Si/O atom in *a*-SiO$_2$.

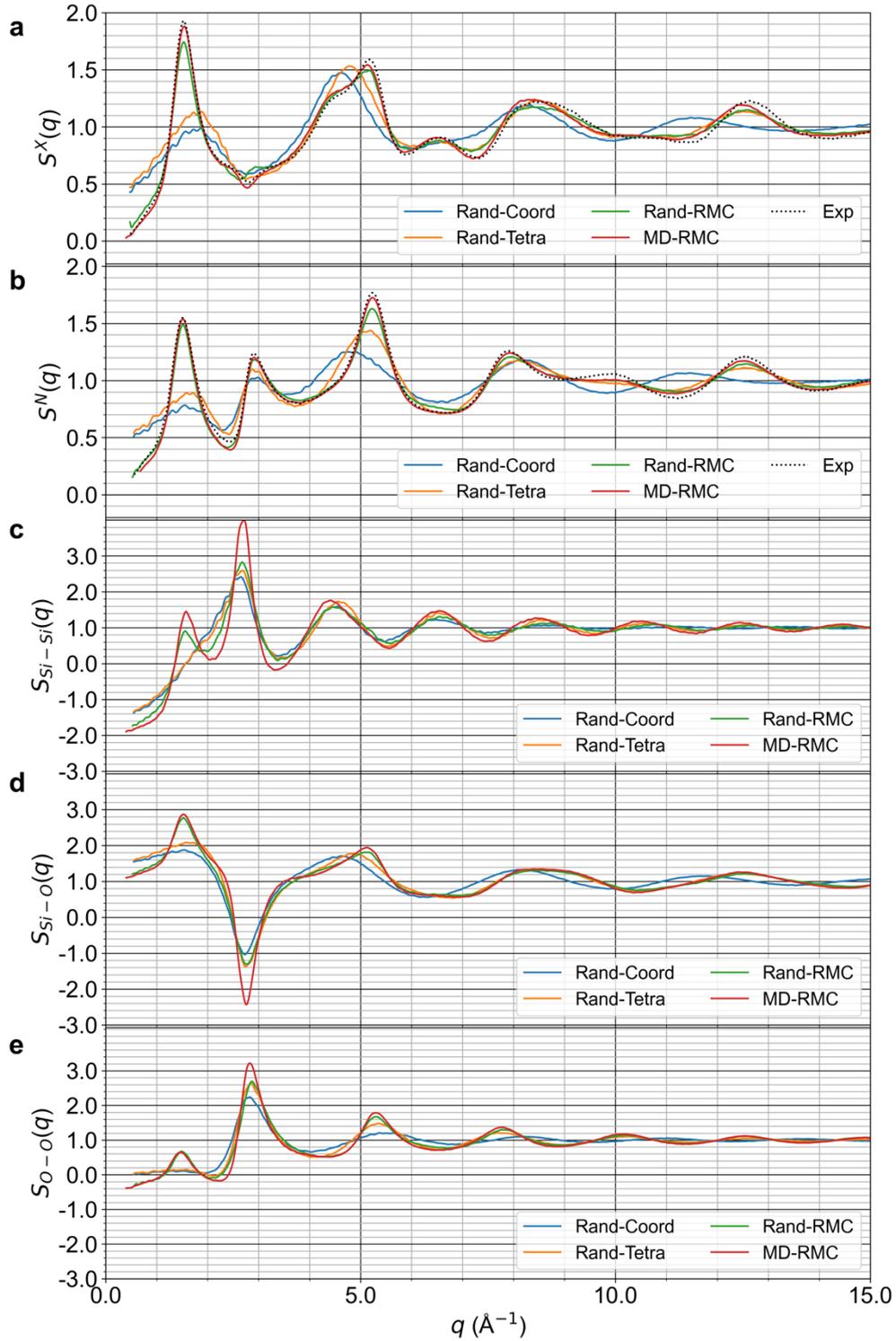

**Supplementary Figure 14. Quantum beam diffraction data for *a*-SiO$_2$ models. a–b** Total structure factors $S(q)$ obtained via X-ray (**a**) and neutron diffractions (**b**). **c–e** Partial structure factors $S_{ij}(q)$ of Si–Si atom pair (**c**), Si–O atom pair (**d**), and O–O atom pair (**e**).

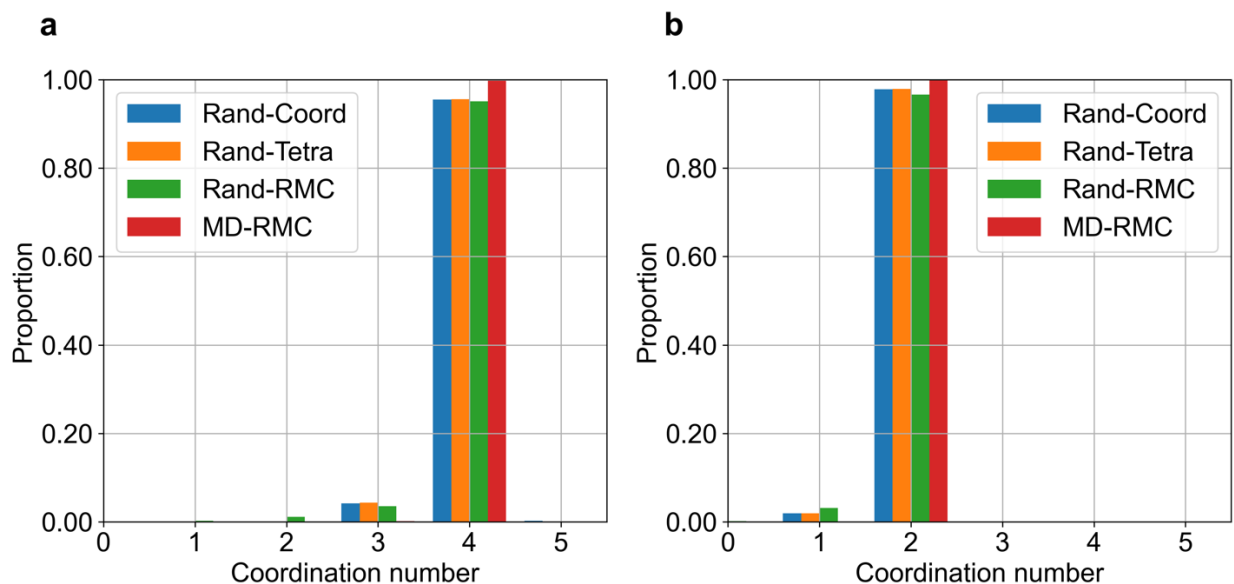

**Supplementary Figure 15. Distribution of coordination numbers in the *a*-SiO$_2$ models.** Coordination numbers of Si atom (**a**) and O atom (**b**).

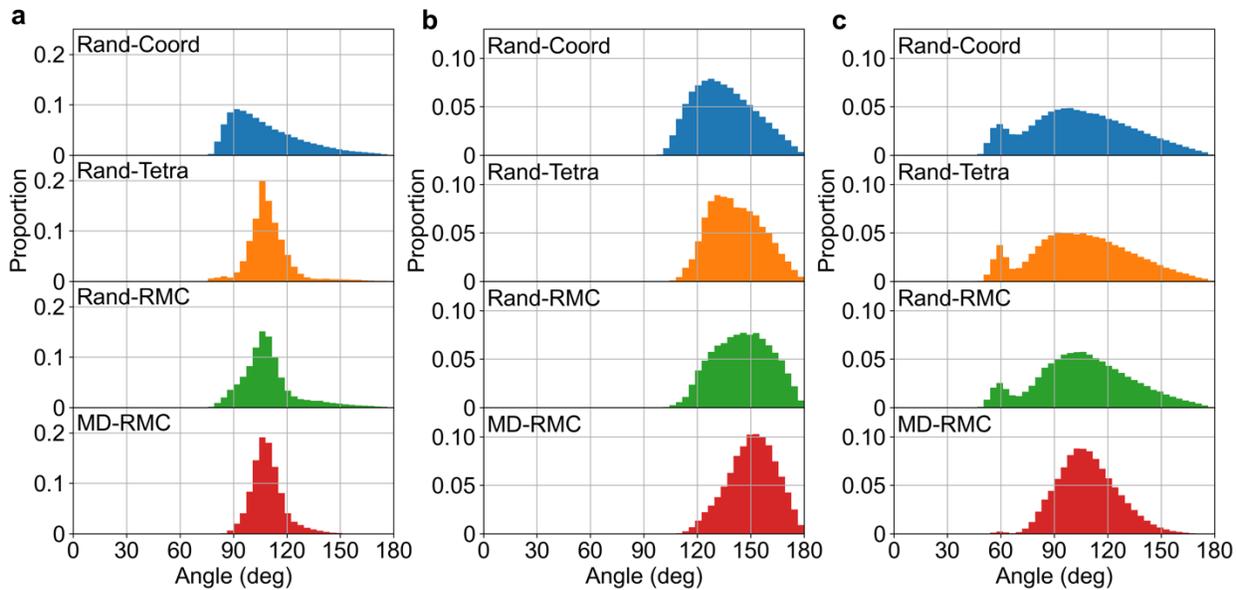

**Supplementary Figure 16. Triplet correlations evaluated by bond angle distribution in the *a*-SiO$_2$ models.** Bond angle distributions of O–Si–O (**a**), Si–O–Si (**b**), and Si–Si–Si (**c**).

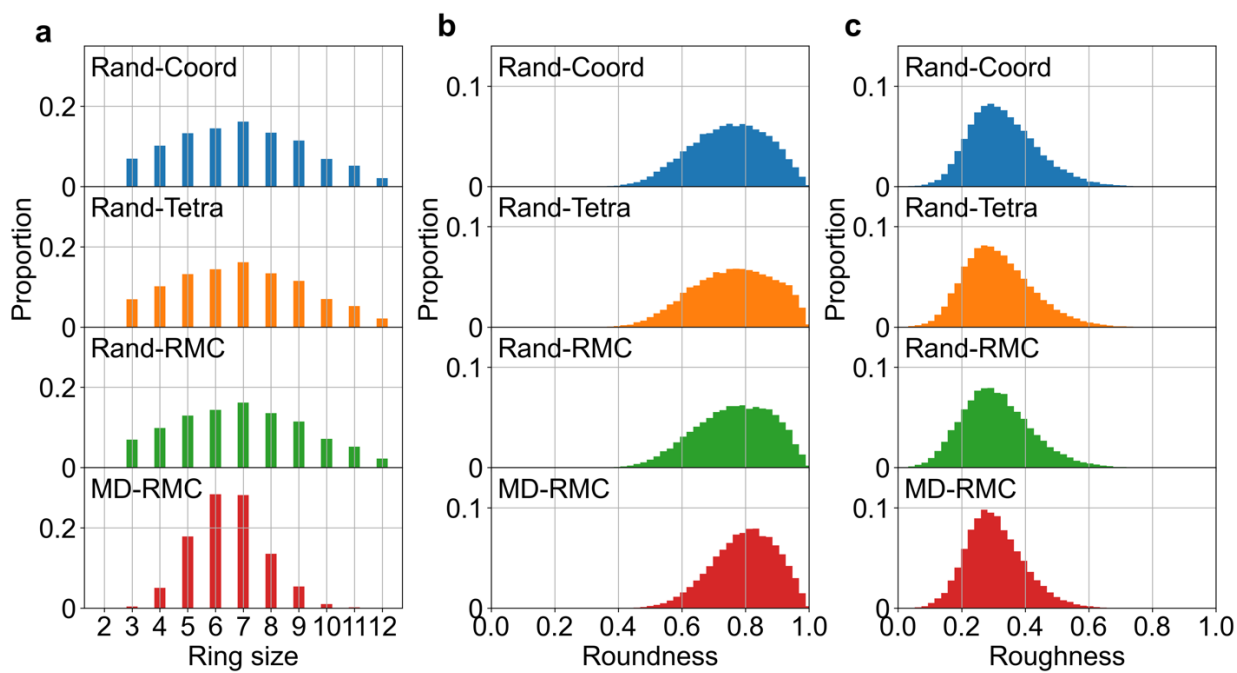

**Supplementary Figure 17. Ring characterizations of the *a*-SiO$_2$ models.** Proportion of the number of rings as a function of ring size (**a**), roundness (**b**), and roughness (**c**).

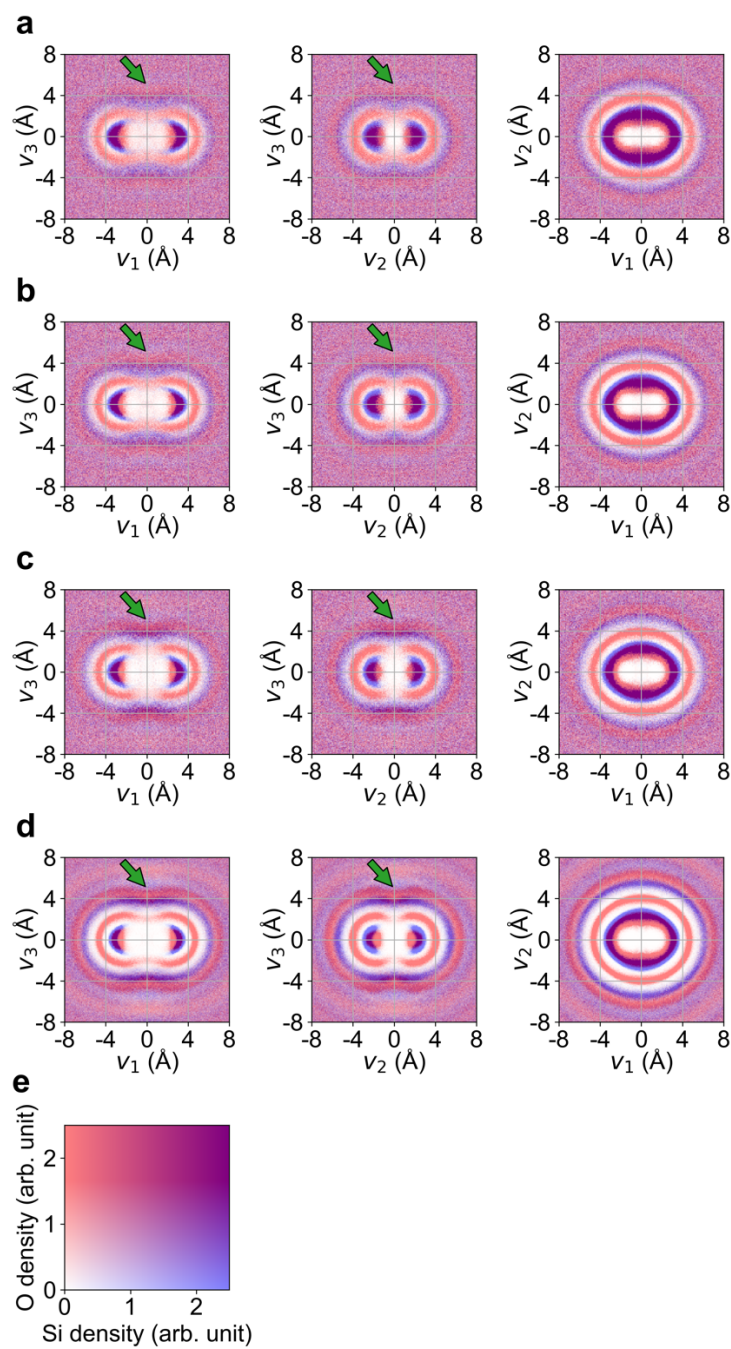

**Supplementary Figure 18. Cross-sectional mappings of spatial correlations computed using only six-fold rings of *a*-SiO$_2$ models**. **a**–**d** Cross-sectional mappings of spatial correlations of Rand-Coord (**a**), Rand-Tetra (**b**), Rand-RMC (**c**), and MD-RMC (**d**). **e** Color indicator, a region with blue/red color, indicates a large density of Si/O atoms. The cross-sectional thickness *t* is 2 Å. Green arrows indicate parallel planes above and below the rings along the $v_1$- and $v_2$-axes.